\documentclass[aps,pra,reprint,showpacs,superscriptaddress]{revtex4-1}
\usepackage{graphics}
\usepackage{graphicx}
\usepackage{epsfig}
\usepackage{amsmath}
\usepackage{amssymb}
\usepackage{amsthm}
\usepackage{calc}
\usepackage{hyperref}
\hypersetup{
    bookmarks=true,%
    citecolor=blue,%
    filecolor=blue,%
    linkcolor=blue,%
    urlcolor=blue
}
\usepackage{color}
\usepackage{lipsum}

\newcommand{\beq}{\begin{equation}}
\newcommand{\eeq}{\end{equation}}
\newcommand{\bea}{\begin{eqnarray}}
\newcommand{\eea}{\end{eqnarray}}

\begin{document}
\title{Coffman-Kundu-Wootters inequality for fermions}
\author{G\'abor S\'arosi}
\author{P\'eter L\'evay}
\affiliation{Department of Theoretical Physics, Institute of
Physics, Budapest University of Technology, H-1521 Budapest,
Hungary}
\date{\today}

\begin{abstract}
We derive an inequality for three fermions with six single particle states which reduces to the sum of the famous Coffman-Kundu-Wootters inequalities when an embedded three qubit system is considered. We identify the quantities which are playing the role of the concurrence, the three-tangle and the invariant $\det \rho_A+\det \rho_B+\det \rho_C$ for this tripartite system. We show that this latter one is almost interchangeable with the von Neumann entropy and conjecture that it measures the entanglement of one fermion with the rest of the system. We prove that the vanishing of the fermionic ``concurrence'' implies that the two particle reduced density matrix is a mixture of separable states. Also the vanishing of this quantitiy is only possible in the GHZ class, where some genuie tripartite entanglement is present and in the separable class. Based on this we conjecture that this ``concurrence'' measures the amount of entanglement between pairs of fermions. We identify the well-known ``spin-flipped'' density 
matrix in the fermionic context as the reduced density matrix of a special particle-hole dual state. We show that in general this dual state is always canonically defined by the Hermitian inner product of the fermionic Fock space and that it can be used to calculate SLOCC covariants. We show that Fierz identities known from the theory of spinors relate SLOCC covariants with reduced density matrix elements of the state and its ``spin-flipped'' dual.
\end{abstract}
\maketitle{}

\tableofcontents

\section{Introduction}

One of the most important ingredients of quantum theory is the phenomenon of entanglement, a feature completely missing from classical mechanics. Entanglement can arise between subsystems of a quantum system: a state is said to be entangled if these subsystems do not have their own pure states. This feature is of key interest in many areas of modern theoretical physics. In the field of quantum computation it is regarded as a resource: entanglement can be used to speed up computational tasks\cite{Nielsen}. It is also an important participant in other areas such as condensed matter physics\cite{Vidalcond}. Even the fabric of spacetime seems to be deeply connected to entanglement as suggested by the Ryu-Takayanagi presciption\cite{Ryu} in AdS/CFT. These wide variety of applications led to the emergence of the vast field of quantum information theory\cite{Nielsen} (QIT) which serves as a common language between different areas of physics with the aim to investigate entanglement. An important 
goal of QIT is to develop methods to 
determine the amount and to classify the type of entanglement a quantum state posesses.

For distinguishable subsystems a wide variety of results is available. The amount of entanglement is usually measured with local unitary (LU) invariants\cite{Bennett,Nielsen} such as the entropy of formation\cite{Wootters2}. Locality is a key concept understood with respect to the subsystems for which entanglement is considered. The LU group then arise from the physically well-motivated idea to measure entanglement: one considers two states equally entangled if they can be obtained from each other with local operations and classical communication\cite{Nielsen} (LOCC). Determining the type of entanglement is also a subtle issue. For example systems with several subsystems can posess multipartite entanglement which turns out to be wildly different and more complicated than bipartite entanglement\cite{Kempe}. One of the most well-known results regarding multipartite entanglement is the result of Coffman, Kundu and Wootters\cite{Coffman}. It is known as the monogamy of 
entanglement: when one considers three qubits A, B and C it turns out that the amount of bipartite entanglement of A 
with the rest of the system is always greater than the sum of the amount of its bipartite entanglement with qubit B and C separately. The difference between the two is the amount of genuie tripartite entanglement and is measured by the so called three-tangle\cite{Coffman}.

 Entanglement between fermions is a slightly different concept: the subsystems under consideration are indistinguishable. However, it is important to keep in mind that one can consider entanglement between distinguishable subsystems of fermionic systems: such as entanglement between particles with high and low momenta. This is the field of study of entanglement between modes\cite{Zanardi,Banuls,Heaney}. On the other hand if one is interested in the entanglement between identical particles as subsystems one is inevitably led to consider entanglement between indistinguishable subsystems. Another reason to study this is that classifying entanglement between fermions turns out to be somewhat easier than the same task for distinguishable subsystems. Since every distinguishable composite system can be embedded into bigger fermionic systems\cite{levvran1,levvran3,DjokPRA} one can gain information about entanglement in distinguishable systems by studying fermionic 
entanglement.
 
 Our main aim in this paper to present an analogue of the equation of Coffman, Kundu and Wootters for three fermions with six single particle states and identify the quantities measuring bipartite and tripartite entanglement. This system is the simplest one that possesses many interesting non-trivial entanglement related features and hence it has already been the subject of many studies\cite{Borland,Djok2,levvran1}. Non-trivial constraints for the important problem of N-representability\cite{Coleman} has been found by Borland and Dennis\cite{Borland}, they have found the necessary and sufficient condition for a one particle reduces density matrix (RDM) to be obtainable from a pure state. The condition is that the eigenvalues of the one particle RDM have to satisfy three equalities and an inequality: these are constraints on unitary invariants of a pure state and hence the result is deeply entanglement-related. This work was much later extended to the general framework of entanglement polytopes by 
Klyachko\cite{Kly1}. Three 
fermions 
with six single particle states is 
also the simplest fermionic system to possess genuie tripartite entanglement\cite{levvran1}. It is in many ways a similar system to the one of three qubits\cite{Dur}. The structure of entanglement classes under stochastic local operations and classical communication\cite{Dur} (SLOCC) - which is basically a coarse graining of LOCC - agree. The system has only one relative invariant under the action of the SLOCC group which is an analougus quantity to the three-tangle for three qubits. This property is due to the fact that the system is a so called prehomogeneous vector space\cite{Kimura,Satokimura}. Three fermions with seven or eight single particle states also form a prehomogeneous vector space with a single SLOCC invariant, however, in the case of nine single particle states there are already four independent continous SLOCC invariants\cite{levsar3} and the physical meaning of these ones is not so clear yet.
 
 The results in this paper are organized as follows. In section \ref{sec:1}. we are studying the system of three fermions with six single particle states. After a short review of the SLOCC classes and the covariant $K$ needed to classify them, in section \ref{sec:CKW}. we derive from a Cauchy-Schwarz inequality that the absolute value of the quartic invariant\cite{levvran1,Satokimura} $\mathcal{D}$ is always less than the value of a spectral based entropy:
 \beq
 6|\mathcal{D}| \leq 3-\text{Tr} \rho^2\leq \frac{3}{2},
 \eeq
 where $\rho$ denotes the one particle reduced density matrix. We define a fermionic \textit{concurrence} as the difference between these two quantities and in section \ref{sec:3qubit}. we show that for three-qubit-like states of three fermions it reduces to the sum of the ususal mixed state squared concurrences of the pairs of qubits: $\mathcal{C}^2_{AB}+\mathcal{C}^2_{AC}+\mathcal{C}^2_{BC}$. In section \ref{sec:spinflipp1}. we show that the ``spin-flipped'' two particle reduced density matrix\cite{Coffman,Hill} - which is used to define the concurrences in the three qubit case but its physical meaning is less transparent - arises as the reduced density matrix of a conjugate particle-hole dual state. In section \ref{sec:vanishingconcurrence}. we analyze states with vanishing fermionic concurrence. We find here that such states are only possible in the GHZ and the fully separable classes\cite{Dur,levvran1}, hence where the presence of any bipartite entanglement is not manifest. We show 
that the spectrum of the one particle RDM and hence 
the von Neumann entropy of these states is a definite function of $|\mathcal{D}|$ or equally a function of the entropy $3-\text{Tr} \rho^2$. In section \ref{sec:bisep}. we find a similar result for states from the biseparable class: the spectrum and the von Neumann entropy is a definite function of the entropy $3-\text{Tr} \rho^2$. In section \ref{sec:entanglemententropy}. we analyze the von Neumann entropy for arbitrary pure states and based on analytic and numeric results we show that the previously studied vanishing concurrence and biseparable states put strict bounds on the relation between the von Neumann entropy and the entropy $3-\text{Tr} \rho^2$. We conclude that the von Neumann entropy is almost interchangeable with $3-\text{Tr} \rho^2$ and hence this latter quantity - which participates in our previously derived inequality - measures entanglement of one fermion with the rest of the system. In section \ref{sec:2particleRDM}. we analyze the two particle RDM. As expected, its non-zero eigenvalues 
agree with the ones of the one particle RDM. We derive that the vanishing of the previously introduced fermionic concurrence implies that the two-particle RDM is a mixture of separable states and hence we conjecture that it measures the amount of bipartite entanglement in a three fermion state. 
 
 In section \ref{sec:2}. we move to considerations for general fermionic systems. We analyze the relation between SLOCC covariants and density matrix elements and find that the bridge between the two is a canonically defined conjugate particle-hole dual state which is a generalization of the ``spin-flipped'' dual for qubits. We show that this state which looks like a charge-conjugate and time-reversed state is always uniquely defined on a fermionic Fock space regardless of any reference to spacetime symmetries. In section \ref{sec:fermandqudits}. we warm up by reminding the reader that a distinguishable composite system can always be considered as a subspace of a fermionic system\cite{levvran3}. In section \ref{sec:extendedSLOCC}. we review the framework of extended SLOCC transformations\cite{levsar2} which is based on the theory of spinors\cite{Chevalley}. In section \ref{sec:bilinprod}. we introduce the invariant bilinear pairing\cite{Chevalley}, known from the theory of spinors, which can be 
used to generate SLOCC 
covariants and invariants for fermions. Finally in section \ref{sec:spinflipp2}. we show that the connection between this bilinear product and the usual Hermitian inner product is made by a conjugate particle-hole dual state. Using this we show that absolute values of SLOCC covariants can be expanded on reduced density matrix elements of the state and its dual and vice-versa. These expansions are known as Fierz identities\cite{Fierz} in the theory of spinors.

\section{Three fermions with six single particle states}
\label{sec:1}

Let
\beq
|P\rangle=\frac{1}{3!}P_{ijk}|e^{ijk}\rangle \in \wedge^3{\mathbb{C}^6}
\eeq
be an unnormalized three fermion state. Here summation for the repeated indices $i,j,k=1...6$ is understood. We use the shorthand notation $|e^{ijk}\rangle=|e^i\rangle \wedge |e^j\rangle \wedge |e^k\rangle$ for a Slater determinant basis of $\wedge^3{\mathbb{C}^6}$ built out of single particle basis vectors $|e^i\rangle$. Equivalently, we could also write $|e^{ijk}\rangle={f^i}^\dagger {f^j}^\dagger {f^k}^\dagger |0\rangle$ in creation operator formalism. 

We briefly review the classification of this system\cite{levvran1,levsar3} under SLOCC. By SLOCC classification we mean the identification of the orbits of the group of invertible SLOCC transformations. The SLOCC group here is $GL(6,\mathbb{C})$. An invertible SLOCC transformation $g\in GL(6,\mathbb{C})$ acts on $|P\rangle$ as
\beq
|P\rangle \mapsto g^* |P\rangle = \frac{1}{3!}(g^*P)_{ijk}|e^{ijk}\rangle \in \wedge^3{\mathbb{C}^6},
\eeq
where 
\beq
(g^*P)_{ijk}={g^{i'}}_i {g^{j'}}_j {g^{k'}}_k P_{i'j'k'},
\eeq
so it is the same invertible transformation of each single particle index.
The SLOCC class of a state can be determined from the calculation of a $6\times 6$ matrix quadratic in the amplitudes:
\beq
\label{eq:Kmatrix}
{K^i}_{j}=\frac{1}{2!3!}\epsilon^{iabcde}P_{jab}P_{cde}.
\eeq
Observe that when we apply the invertible SLOCC transformation $g\in GL(6,\mathbb{C})$ on $|P\rangle$ the matrix $K$ transforms as 
\beq
K \mapsto (\det g) g^{-1}Kg.
\eeq
Now it is manifest that the rank of $K$ is invariant under an invertible SLOCC transformation. It turns out that this rank determines the SLOCC class of $P$ uniquely. We review these classes in Table \ref{tab:1}.

\begin{table}[h!]
\centering
\begin{tabular}{ccc}
\hline \hline
Type & Canonical form of $P$ & Rank $K$ \\
 \hline 
Sep & $|e^{123}\rangle$ & 0 \\
Bisep & $|e^{123}\rangle +|e^{156}\rangle$ & 1  \\
W & $|e^{126}\rangle+|e^{423}\rangle+|e^{153}\rangle$ & 3  \\
GHZ & $|e^{123}\rangle+|e^{456}\rangle$ & 6  \\  \hline \hline
\end{tabular}
\caption{Entanglement classes of three fermions with six single
particle states with their canonical forms, and the rank of the matrix $K$.}
\label{tab:1}
\end{table}

It is also easy to see that the trace of powers of $K$ are relative invariants under invertible SLOCC transformation. The term relative invariant refers to a quantity wich picks up only a one dimensional character when transformed: in this case a power of the determinant of a SLOCC transformation. These expressions are polynomials of the amplitudes $P_{ijk}$ and we call them polynomial invariants. It turns out that in the case of three fermions with six single particle states we have only one algebraically independent such invariant\cite{levvran1,Satokimura,Djok2}. Now $\text{Tr}K=0$ for every state while the trace of the second power gives the aforementioned quartic SLOCC relative invariant:
\beq
\label{eq:quartic}
\mathcal{D}(P)=\frac{1}{6}\text{Tr}K^2.
\eeq
Note that the condition $\mathcal{D}(P)\neq 0$ singles out the GHZ class of Table \ref{tab:1}. It is also an important property that the square of $K$ is always proportional to the identity:
\beq
\label{eq:Ksquare}
K^2=\mathcal{D}(P) I.
\eeq
This can be easily checked by plugging in the canonical GHZ state of Table \ref{tab:1}. and using the relative covariance of $K^2$.

Now define the one particle reduced density matrix (RDM) as
\beq
{\rho_{i}}^j=\frac{1}{2} P_{inm}\bar P^{jnm}.
\eeq
This is a covariant quantity under local unitary transformations (corresponding to invertible LOCC) but not under local invertible transformations (corresponding to invertible SLOCC). Note that we have adopted the convention that a complex conjugation raises an index, however indices contracted this way only stay invariant under unitary transformations. Note also that we have adopted L\"owdin normalization\cite{Coleman}: $\text{Tr}\rho = 3||P||^2$. The entropies $\text{Tr}\rho^n$, $n\in \mathbb{N}$ are local unitary invariants of $|P\rangle$. Equivalently one can look at the eigenvalues $\lambda_i$, $i=1...6$ of $\rho$ which are also unitary invariants. We recall the classical result of Borland and Dennis\cite{Borland} that these eigenvalues satisfy the (in)equalities:
\beq
\label{eq:BorlandDennis}
\begin{aligned}
 \lambda_1 + \lambda_6=1, &&
 \lambda_2 + \lambda_5=1, \\
 \lambda_3+\lambda_4 =1, &&
 \lambda_5 +\lambda_6 \geq \lambda_4,
\end{aligned}
\eeq
where the eigenvalues are ordered as $\lambda_1\geq \lambda_2 \geq ... \geq \lambda_6$. An important fact to be used later is that in general there are only three of these eigenvalues are independent.

\subsection{The CKW inequality}
\label{sec:CKW}

Using the usual identity for Levi-Civita symbols
\beq
\epsilon^{i_1i_2i_3i_4i_5i_6}\epsilon_{j_1j_2j_3j_4j_5j_6}={\delta^{i_1}}_{[j_1}{\delta^{i_2}}_{j_2}{\delta^{i_3}}_{j_3}{\delta^{i_4}}_{j_4}{\delta^{i_5}}_{j_5}{\delta^{i_6}}_{j_6]},
\eeq
(the brackets denote antisymmetrization) one can easily calculate that the following equality holds
\beq
\label{eq:KKdagger}
\text{Tr}(K K^\dagger) = \frac{1}{3}((\text{Tr}\rho)^2-3\text{Tr}\rho^2).
\eeq
Note that on the left hand side we have a product of a SLOCC covariant quantity with its conjugate while on the right hand side we have a product of LOCC covariant quantities. We will later see in section \ref{sec:spinflipp2}. that this is nothing else but a Fierz identity known from the theory of spinors. These type of relations will be of much interest for us later on. For normalized states we have $\text{Tr}\rho=3$ and hence $\text{Tr}(K K^\dagger) =3-\text{Tr}\rho^2$. We see that $\text{Tr}(K K^\dagger)$ measures the distance of $\text{Tr}\rho^2$ from its value on a separable (single Slater determinant) state.

Now since for any matrix $M$ the matrix $M M^\dagger$ is positive we have the following inequality:
\beq
0\leq \text{Tr}\left((K-\lambda K^\dagger)(K^\dagger -\bar \lambda K)\right),
\eeq
for any $\lambda \in \mathbb{C}$. It is easy to see that the righ hand side has minimum at $\lambda=\frac{\text{Tr}K^2}{|\text{Tr}K^2|}$. Plugging this back we obtain the inequality
\beq
|\text{Tr}(K^2)|\leq \text{Tr}(KK^\dagger).
\eeq
Note that this is just the Cauchy-Schwarz inequality for the Hilbert-Schmidt product of $K$ with $K^\dagger$. This gives a lower bound on $\text{Tr}(KK^\dagger)$ in terms of the quartic invariant of e.q. \eqref{eq:quartic}. On the other hand for normalized states one has $\text{Tr}KK^\dagger = \text{Tr}\rho(I-\rho)$. Thus we also have an upper bound for $\text{Tr}KK^\dagger$ due to the Cauchy-Schwarz inequality between $\rho$ and $I-\rho$:
\beq
\text{Tr}KK^\dagger = \text{Tr}\rho(I-\rho)\leq \sqrt{\text{Tr}\rho^2 \text{Tr}(I-\rho)^2}.
\eeq
But we know from the classical paper of Borland and Dennis\cite{Borland} (see e.q. \eqref{eq:BorlandDennis}) that the eigenvalues of $\rho$ come in pairs and each pair sums to one. This implies that $\text{Tr}\rho^2=\text{Tr}(I-\rho)^2=\sum_{i=1}^3(\lambda_i^2+(1-\lambda_i)^2)$. Using this and \eqref{eq:KKdagger} we arrive at
\beq
\text{Tr}KK^\dagger \leq \text{Tr}\rho^2=3-\text{Tr}KK^\dagger,
\eeq
or after rearrangement
\beq
\text{Tr}KK^\dagger \leq \frac{3}{2}.
\eeq
Note that this inequality is saturated if and only if $\rho = I-\rho$ and hence $\rho=\frac{1}{2}I$ which means that the one particle RDM is maximaly mixed.

We define the non-negative quantity:
\beq
\label{eq:fermconc}
Con(P)=\text{Tr}(KK^\dagger)-|\text{Tr}(K^2)|,
\eeq
which satisfies
\beq
0 \leq Con(P) \leq \text{Tr}(KK^\dagger)\leq \frac{3}{2}.
\eeq
As we will show in the next subsection $Con(P)$ is the sum of the concurrences for embedded three qubit states.

Note that the concurrence vanishes for a state $P$ if and only if it satisfies $K=\frac{\text{Tr}K^2}{|\text{Tr}K^2|}K^\dagger$. Here $\frac{\text{Tr}K^2}{|\text{Tr}K^2|}$ is just the phase of the quartic invariant.

\subsection{Embedded three qubit states}
\label{sec:3qubit}

Consider the three qubit state 
\beq
|\psi\rangle = \sum_{ijk=0}^1 \psi_{ijk}|ijk\rangle \;\;\in \mathbb{C}^2\otimes \mathbb{C}^2 \otimes \mathbb{C}^2.
\eeq
One can embed this state into the space of three fermions with six single particle states in several ways and realize three qubit SLOCC transformations as special fermionic SLOCC transformations\cite{levvran1,levsar2}. A convenient choice is 
\beq
\label{eq:3qubitembedding}
\begin{aligned}
P^\psi_{123}=\psi_{000}, && P^\psi_{12\bar 3}=\psi_{001}, && P^\psi_{1\bar 23}=\psi_{010},\\ P^\psi_{1 \bar 2\bar 3}=\psi_{011}, &&
P^\psi_{\bar 123}=\psi_{100}, && P^\psi_{\bar 12\bar 3}=\psi_{101}, 
\\ P^\psi_{\bar 1\bar 23}=\psi_{110}, && P^\psi_{\bar 1 \bar 2\bar 3}=\psi_{111},
\end{aligned}
\eeq
with all other (independent) amplitudes being zero. Here we have introduced the relabelling of indices $\lbrace 1,2,3,4,5,6\rbrace \leftrightarrow \lbrace 1,2,3,\bar 1,\bar 2,\bar 3\rbrace$. With this choice a three qubit SLOCC transformation $g^{(1)}\otimes g^{(2)} \otimes g^{(3)} \in GL(2,\mathbb{C})^{\times 3}$ takes the form
\beq
g=
\left(
\begin{array}{ccc}
g^{(1)} & & \\
 & g^{(2)} & \\
 & & g^{(3)} 
\end{array}
\right)
\eeq
where the order of indices in the matrix $g\in GL(6,\mathbb{C})$ is $1,\bar 1,2,\bar 2,3,\bar 3$. One can think of this embedding as selecting the subspace of the single occupancy states\cite{DjokPRA,levsar2} of three spin $\frac{1}{2}$ fermions which can occupy three nodes. For example $|e^{12\bar 3}\rangle$ represents a state where the first two nodes are occupied by spin-up fermions while the third one is occupied by a spin-down fermion. 

Let us denote the $K$ matrix defined in \eqref{eq:Kmatrix} for the state $P^\psi$ with $K_\psi$. It is well known that under this embedding the quartic invariant of three fermions reduces to Cayley's hyperdeterminant\cite{levvran1,Cayley}:
\beq
\begin{aligned}
\frac{1}{6}\text{Tr}K_\psi^2 &= \psi_{000}^2 \psi_{111}^2+\psi_{001}^2 \psi_{110}^2+\psi_{010}^2 \psi_{101}^2+\psi_{100}^2  \psi_{011}^2\\ &-2(\psi_{000}\psi_{001}\psi_{100}\psi_{111} +\psi_{000}\psi_{010}\psi_{101}\psi_{111}\\  &+\psi_{000}\psi_{100}\psi_{011}\psi_{111} +\psi_{001}\psi_{010}\psi_{101}\psi_{110} \\ &+\psi_{001}\psi_{100}\psi_{011}\psi_{110}+\psi_{010}\psi_{100}\psi_{011}\psi_{101}) \\ &+4(\psi_{000}\psi_{011}\psi_{101}\psi_{110}+\psi_{001}\psi_{010}\psi_{100}\psi_{111}).
\end{aligned}
\eeq
The three-tangle $\tau_{ABC}$ for three qubits is defined to be four times the absolute value of Cayley's hyperdeterminant\cite{Coffman}:
\beq
\label{eq:tangle}
\tau_{ABC}=\frac{2}{3}|\text{Tr}K_\psi^2|.
\eeq
On the other hand the fermionic one particle reduced density matrix for the state $P^\psi$ has the form
\beq
\rho=
\left(
\begin{array}{ccc}
\rho_A & & \\
 & \rho_B & \\
 & & \rho_C 
\end{array}
\right),
\eeq
where $\rho_A,\rho_B,\rho_C$ are one particle RDMs of the three qubit state $|\psi\rangle $. For normalized three qubit states we have $\text{Tr}\rho_{A,B,C}=1$ and $\text{Tr}\rho=3$. Using that for $2\times 2$ matrices one has the identity $2\det \rho_{A,B,C} = (\text{Tr}\rho_{A,B,C})^2-\text{Tr}({\rho_{A,B,C}}^2)$, one can easily show that e.q. \eqref{eq:KKdagger} reduces to
\beq
\label{eq:KKdagger2}
\text{Tr} (K_\psi K_\psi^\dagger)=2(\det \rho_A+\det \rho_B+\det \rho_C ).
\eeq
Now recall the Coffman-Kundu-Wootters equations\cite{Coffman} for three qubits which read as
\beq
\begin{aligned}
4\det \rho_A=\tau_{ABC}+\mathcal{C}^2_{AB}+\mathcal{C}^2_{AC},\\
4\det \rho_B=\tau_{ABC}+\mathcal{C}^2_{AB}+\mathcal{C}^2_{BC},\\
4\det \rho_C=\tau_{ABC}+\mathcal{C}^2_{CA}+\mathcal{C}^2_{CB},
\end{aligned}
\eeq
where $\mathcal{C}^2_{AB},\mathcal{C}^2_{AC},\mathcal{C}^2_{BC}$ are the concurrences\cite{Hill,Coffman,Wootters} between the qubits $AB$, $AC$ and $BC$ respectively. Adding up these equations and using equations \eqref{eq:tangle} and \eqref{eq:KKdagger2} one arrives at
\beq
\mathcal{C}^2_{AB}+\mathcal{C}^2_{AC}+\mathcal{C}^2_{BC}+|\text{Tr}K_\psi^2|=\text{Tr} (K_\psi K_\psi^\dagger).
\eeq
Compare with the definition \eqref{eq:fermconc} to arrive at
\beq
Con(P^\psi)=\mathcal{C}^2_{AB}+\mathcal{C}^2_{AC}+\mathcal{C}^2_{BC}.
\eeq
Note that these equations written with the use of $K_\psi$ are invariant under the action of the fermionic LOCC group $U(6)$. Hence we could choose any other isometric embedding of three qubits into three fermions with six single particle states and we would have obtained the same results. 

\subsection{The ``spin-flipped'' density matrix}
\label{sec:spinflipp1}

For three qubit states the concurrence is defined via the so called ``spin-flipped'' density matrix\cite{Hill} $\tilde \rho^{AB}$. It is defined to be 
\beq
\tilde \rho^{AB} = (\sigma_y \otimes \sigma_y) \bar{ \rho}^{AB}(\sigma_y \otimes \sigma_y),
\eeq
where $\sigma_y$ is the second Pauli matrix and $\rho^{AB}$ is the two-particle RDM of qubits $A$ and $B$:
\beq
\rho^{AB}_{ij|kl}=\sum_{n=0}^1 \psi_{ijn}\bar \psi_{kln}.
\eeq
The matrix $\rho^{AB}\tilde \rho^{AB}$ has non-negative real eigenvalues. Let these be $\lambda_1^2,\lambda_2^2,\lambda_3^2,\lambda_4^2$ in decreasing order. The concurrence between qubit $A$ and $B$ is defined to be\cite{Hill,Wootters}
\beq
\mathcal{C}_{AB}=\text{max}\lbrace \lambda_1-\lambda_2-\lambda_3-\lambda_4,0\rbrace.
\eeq
Here we show that the ``spin-flipped'' density matrix has a clear physical meaning in the fermionic context namely it arises from the RDM of the complex conjugate of the particle-hole dual state. Later on in section \ref{sec:spinflipp2}. we will show that once the Hermitian inner product is fixed this dual state is allway canonically and unambigously defined for any state in the fermionic Fock space. Here we define the dual state for three fermions with six single particle states as
\beq
|\tilde P \rangle =\frac{1}{3!}\tilde P_{ijk}|e^{ijk}\rangle ,
\eeq
with
\beq
\label{eq:Pdual}
\tilde P_{ijk} = \frac{1}{3!}\epsilon_{ijklmn}\bar P^{lmn}.
\eeq
We will see later that it is important that the map $|P\rangle \mapsto |\tilde P\rangle$ is antilinear. It is not difficult to see that for the embedded three qubit state $P^\psi$, the dual $\tilde P^\psi$ can be obtained with our embedding \eqref{eq:3qubitembedding} from the three qubit state $|\tilde \psi\rangle$ with coefficients:
\beq
\tilde \psi_{ijk}=\sum_{i'j'k'=0}^1 \varepsilon_{ii'}\varepsilon_{jj'}\varepsilon_{kk'}\bar \psi_{i'j'k'},
\eeq
where $\varepsilon$ is the totaly antisymmetric $2\times 2$ matrix or $\varepsilon=i\sigma_y$. Now using $\varepsilon^T\varepsilon=1$ it is straightforward to see that the ``spin-flipped'' density matrix is just the two particle RDM of qubits $A$ and $B$ for the state $|\tilde \psi\rangle$. It is not difficult to see how the two particle RDMs of three qubits sit inside the fermionic two particle RDMs. Let $\rho^{(2)}$ denote the fermionic two particle RDM with coefficients 
\beq
{\rho^{(2)}_{ij}}^{kl}=\frac{1}{2}P_{ijn}\bar P^{kln}.
\eeq
We denote by $\tilde \rho^{(2)}$ the two particle RDM of the dual state $|\tilde P\rangle$. For an embedded state $P^\psi$ we have
\beq
\begin{aligned}
 {\rho^{(2)}_{12}}^{12}=\frac{1}{2}\rho^{AB}_{00|00}, && {\rho^{(2)}_{12}}^{1\bar 2}=\frac{1}{2}\rho^{AB}_{00|01}, && {\rho^{(2)}_{12}}^{\bar 1 \bar 2}=\frac{1}{2}\rho^{AB}_{00|11}, \\
 {\rho^{(2)}_{1\bar 2}}^{12}=\frac{1}{2}\rho^{AB}_{01|00}, && ..., && {\rho^{(2)}_{\bar1 \bar2}}^{\bar 1\bar 2}=\frac{1}{2}\rho^{AB}_{11|11}, \\
 {\rho^{(2)}_{13}}^{13}=\frac{1}{2}\rho^{AC}_{00|00}, && ..., && {\rho^{(2)}_{\bar 1\bar 3}}^{\bar1 \bar 3}=\frac{1}{2}\rho^{AC}_{11|11},
\end{aligned}
\eeq
and so on. One observes the scheme that the indices $1,2,3$ correspond to qubits $A,B,C$ respectively. Cases where index pairs contain the same number multiple times like ${\rho^{(2)}_{2\bar 2}}^{12}$ or where the lower and upper indices contain different numbers like ${\rho^{(2)}_{12}}^{13}$ give zero. Indices without a bar correspond to the corresponding qubit in $|0\rangle$ state while with a bar to the qubit in $|1\rangle$ state. The ``spin flipped'' density matrix sits in $\tilde \rho^{(2)}$ in exactly the same way. Note that in general we have
\beq
\text{Tr}(KK^\dagger)=\text{Tr}(\rho^{(2)} \tilde \rho^{(2)}),
\eeq
or for the embedded three qubit state
\beq
\text{Tr}(K_\psi K^\dagger_\psi)=\text{Tr}(\rho^{AB} \tilde \rho^{AB}+\rho^{AC} \tilde \rho^{AC}+\rho^{BC} \tilde \rho^{BC}),
\eeq
which estabilishes a connection with the unnormalized form of the CKW inequality\cite{Coffman}:
\beq
\begin{aligned}
Con(P^\psi) &=\mathcal{C}^2_{AB}+\mathcal{C}^2_{AC}+\mathcal{C}^2_{BC} \\ &\leq \text{Tr}(K_\psi K^\dagger_\psi)\\
&=\text{Tr}(\rho^{AB} \tilde \rho^{AB}+\rho^{AC} \tilde \rho^{AC}+\rho^{BC} \tilde \rho^{BC}).
\end{aligned}
\eeq

It is not difficult to show that the one particle RDM of the dual state \eqref{eq:Pdual} is just $\tilde \rho = \frac{\text{Tr}\rho}{3}I-\rho$, or for normalized states
\beq
\label{eq:dualhole}
\tilde \rho =I-\rho.
\eeq
Since the eigenvalues of a fermionic one particle RDM are interpreted as occupation numbers this equation reinforces the particle-hole duality picture behind the ``spin-flipped'' state. Later on, in section \ref{sec:spinflipp2}., we will show that this equation holds for an arbitary number of fermions with arbitary many single particle states.

\subsection{Vanishing concurrence}
\label{sec:vanishingconcurrence}

Recall that the condition for the vanishing of the quantity $Con(P)$ defined in e.q. \eqref{eq:fermconc} was that 
\beq
K=e^{i\varphi}K^\dagger,
\eeq
where we have introduced for the phase of the quartic invariant $e^{i\varphi}=\frac{\text{Tr}K^2}{|\text{Tr}K^2|}$. This equation has some interesting consequences. With a little bit of work one can show that although the matrices $KK^\dagger$ and $K^\dagger K$ do depend on the two particle RDM, the anticommutator $\lbrace K,K^\dagger\rbrace$ is only a function of the one particle RDM:
\begin{widetext}
\beq
\begin{aligned}
 {(\lbrace K ,K^\dagger\rbrace)^i}_k =
\frac{1}{3}{\delta^i}_k  & \left((\text{Tr}\rho)^2 -3\text{Tr}\rho^2\right)-4{\left(\rho(\frac{\text{Tr}\rho}{3}I-\rho)\right)_k}^i
\end{aligned}
\eeq
\end{widetext}
Using equation \eqref{eq:KKdagger} and that the one particle RDM of the dual state \eqref{eq:Pdual} is just $\tilde \rho = \frac{\text{Tr}\rho}{3}I-\rho$ one can write this as
\beq
\label{eq:rorotilde}
{(\lbrace K,K^\dagger\rbrace)^i}_k={\delta^i}_k\text{Tr}(KK^\dagger)-4{\left(\rho \tilde \rho\right)_k}^i
\eeq
Now using the condition of vanishing concurrence we arrive at
\beq
2 e^{-i\varphi}{\left(K^2\right)^i}_k={\delta^i}_k|\text{Tr}K^2|-4{\left(\rho \tilde \rho\right)_k}^i
\eeq
Now it is well known that for any state the square of $K$ is proportional to the identity matrix: $K^2=\frac{1}{6}(\text{Tr}K^2)I$ (see e.q. \eqref{eq:Ksquare}). Plugging this in we arrive at
\beq
{\left(\rho \tilde \rho\right)_k}^i=|\frac{1}{6}\text{Tr}K^2|{\delta^i}_k = |\mathcal{D}(P)|{\delta^i}_k. 
\eeq
Now suppose that $P$ is normalized. Then we have $\text{Tr}\rho =3$ and as a consequence $\rho \tilde \rho = \rho(I-\rho)$. Using this we see that all the $\lambda_i$ eigenvalues of $\rho$ satisfy the equations
\beq
\lambda_i(1-\lambda_i)=|\mathcal{D}(P)|.
\eeq
The two roots of this equation are
\beq
\begin{aligned}
\lambda^*, && 1-\lambda^*, &&\text{where}\;\;
\lambda^*=\frac{1}{2}\left(1+\sqrt{1-4|\mathcal{D}(P)|} \right).
\end{aligned}
\eeq
Now $\sum_i\lambda_i=3$ implies that three of the eigenvalues are $\lambda^*$ and the other three are $1-\lambda^*$. In particular for these states the von Neumann entropy is always expressable as a function of the quartic invariant:
\beq
\label{eq:zeroconentropy}
\begin{aligned}
S_N &=-\text{Tr}\frac{\rho}{3} \log \frac{\rho}{3} \\ &= 3h(\lambda^*)\\ &=3h\left(\frac{1}{2}\left(1+\sqrt{1-2/3\text{Tr}KK^\dagger} \right)\right),
\end{aligned}
\eeq
where 
\beq
\label{eq:binentropy}
h(x)=-\frac{1}{3}\left( x\log \frac{x}{3}+(1-x)\log(\frac{1-x}{3})\right).
\eeq
In the last line of e.q. \eqref{eq:zeroconentropy} we have used $\text{Tr}KK^\dagger=6|\mathcal{D}(P)|$ which is valid for states with zero concurrence. The $1/3$ factors in the binary entropy function appear to restore the probability normalization of the eigenvalues from the $\text{Tr}\rho=3$ normalization. Observe that if $\mathcal{D}(P)=0$ then three of the eigenvalues are $1$ and the other three are 0 hence the original state is in the separable class. This shows that the vanishing of $Con(P)$ is only possible in the GHZ and the separable classes where the presence of two-particle entanglement is not manifest.

\subsection{Biseparable class}
\label{sec:bisep}

Recall from Table \ref{tab:1}. that in the biseparable class one has rank$K=1$. This means that there exist two vectors $u$ and $v$ in $\mathbb{C}^6$ such that $K$ is a dyad:
\beq
{K^i}_j=u^i\bar v_j,
\eeq
or $K=u v^\dagger$ in index-free notation. From $\text{Tr}K=0$ we see that $v^\dagger u=\bar v_i u^i=0$. A short calculation gives 
\beq
\lbrace K,K^\dagger \rbrace = |v|^2 u u^\dagger + |u|^2 v v^\dagger.
\eeq
Since the projections $u u^\dagger$ and $v v^\dagger$ are orthogonal this is already a diagonal decomposition of $\lbrace K,K^\dagger \rbrace$. We observe that $\text{Tr}KK^\dagger=|u|^2|v|^2$. Using \eqref{eq:rorotilde} we write
\beq
|v|^2 u u^\dagger + |u|^2 v v^\dagger-|u|^2|v|^2 I=-4\rho(I-\rho).
\eeq
The eigenvectors of the left hand side are $u$, $v$ and four arbitrary vectors that are orthogonal to $u$ and $v$. The eigenvectors of the right hand side are just the eigenvectors of $\rho$ itself. Multiplying the equation with $u$ or $v$ gives the equation
\beq
0=\lambda(1-\lambda).
\eeq
We conclude that we have $\lambda_6=0$ and $\lambda_1=1$ as the eigenvalues corresponding to $u$ and $v$. Multiplying with any eigenvector of $\rho$ orthogonal to $u$ and $v$ we get the equation
\beq
\lambda(1-\lambda)=\frac{|u|^2|v|^2}{4}=\frac{\text{Tr}KK^\dagger}{4}.
\eeq
We conclude that the remaining eigenvalues are
\beq
\begin{aligned}
\lambda_2 &=\lambda_3= \frac{1+\sqrt{1-\text{Tr}KK^\dagger}}{2}\\
\lambda_4 &=\lambda_5=1-\lambda_3.
\end{aligned}
\eeq
For the von Neumann entropy note that for the binary entropy function defined in \eqref{eq:binentropy} we have $\lim_{x\rightarrow 0} h(x)= \frac{\log 3}{3}$. Using this we arrive at
\beq
\label{eq:bisepentropy}
S_N=-\text{Tr}\frac{\rho}{3} \log \frac{\rho}{3} = \frac{\log 3}{3}+2 h\left(\frac{1+\sqrt{1-\text{Tr}KK^\dagger}}{2}\right), 
\eeq
for biseparable states.

\subsection{Entanglement entropy of arbitrary states}
\label{sec:entanglemententropy}

We have seen that for biseparable states and for states with vanishing $Con(P)$ we have the entanglement entropy $S_N=-\text{Tr}\frac{\rho}{3} \log \frac{\rho}{3}$ as a definite function of the quantity $\text{Tr}KK^\dagger$. For general states the two are not a function of each other but they are almost interchangeable.

To see this first we prove that the von Neumann entropy of zero concurrence states is an upper bound to the entropy of all states. Denote the eigenvalues of the matrix $\frac{1}{4}((\text{Tr}KK^\dagger) I-\lbrace K,K^\dagger\rbrace)$ with $\mu_i$. From \eqref{eq:rorotilde} one sees that $\mu_i=\lambda_i(1-\lambda_i)$ holds for the $\lambda_i$ eigenvalues of $\rho$. Hence for the von Neumann entropy we have
\beq
S_N=-\sum_{i=1}^6\frac{\lambda_i}{3} \log \frac{\lambda_i}{3}  =\sum_{i=1}^3 h\left(\frac{1}{2}\left(1+\sqrt{1-4 \mu_i} \right)\right)
\eeq
Recall the classical Borland-Dennis results on the spectrum of the on particle RDM: the spectrum consists of three pairs of eigenvalues and every pair sum to one (see e.q. \eqref{eq:BorlandDennis}). Using this we have $\text{Tr}KK^\dagger=\sum_{i=1}^6\mu_i=2\sum_{i=1}^3\mu_i$. Now it is easy to see that the function
\beq
h\left(\frac{1}{2}\left(1+\sqrt{1-2/3 x} \right)\right)
\eeq
is concave, hence
\begin{widetext}
\beq
\begin{aligned}
3h \left(\frac{1}{2}\left(1+\sqrt{1-2/3\text{Tr}KK^\dagger} \right)\right) &=
 3h\left(\frac{1}{2}\left(1+\sqrt{1-4/3 \sum_{i=1}^3\mu_i} \right)\right) \\
 & \geq \sum_{i=1}^3 h\left(\frac{1}{2}\left(1+\sqrt{1-4 \mu_i} \right)\right) \\
 &=S_N,
\end{aligned}
\eeq
\end{widetext}
and hence the claim follows.

\begin{figure}[h!]
\centering
\includegraphics[width=0.45\textwidth]{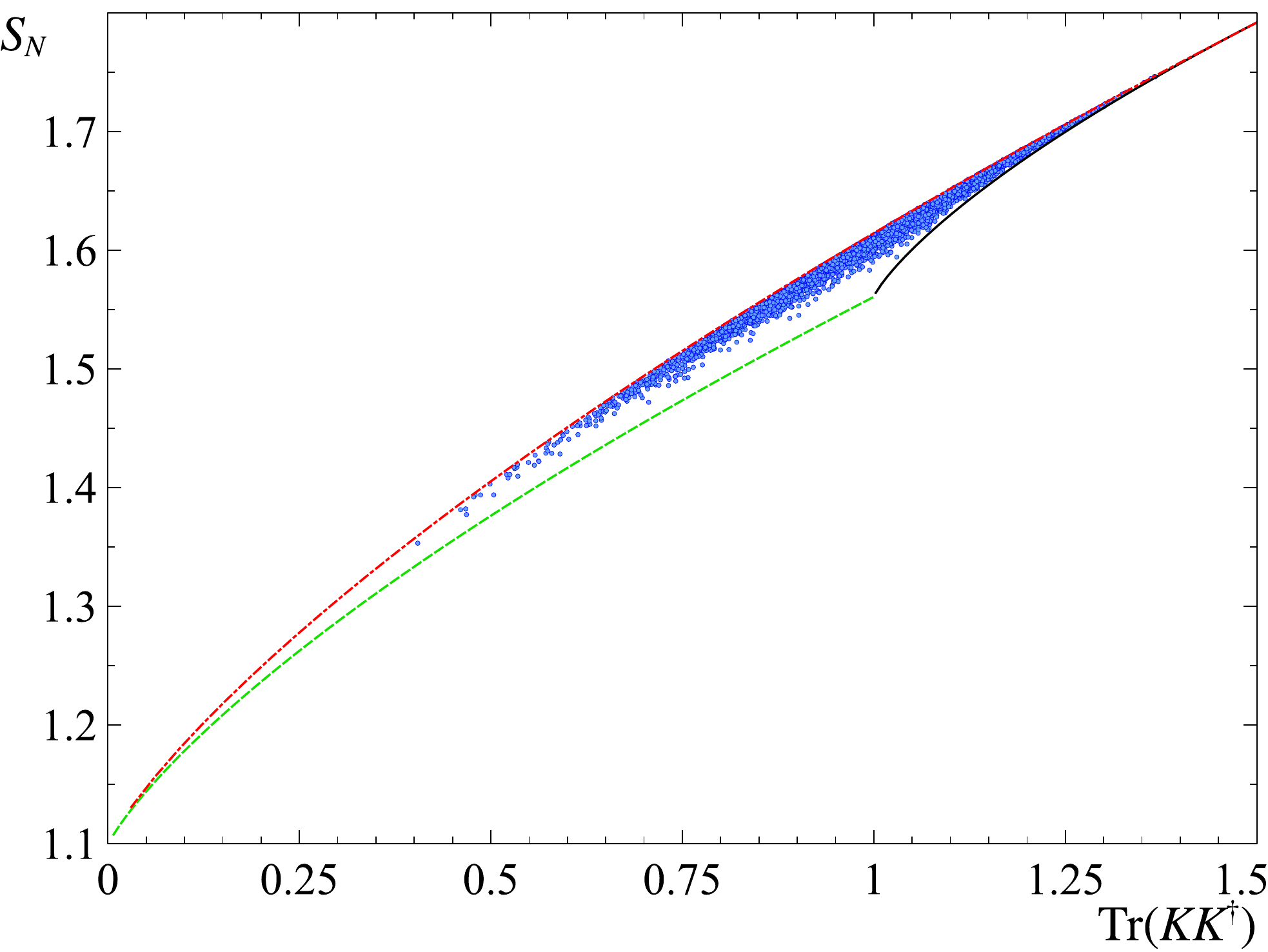}
\caption{The distribution of $\text{Tr}KK^\dagger$ and the entanglement entropy $S=-\text{Tr}\frac{\rho}{3} \log \frac{\rho}{3}$ of random generated states of three fermions with six single particle states. The blue circles are 5000 random generated states in the GHZ class. The dashed-dotted red line is the function \eqref{eq:zeroconentropy} valid for states with zero concurrence while the dashed green line is the function \eqref{eq:bisepentropy} valid for biseparable states. The solid black line is the entropy calculated from \eqref{eq:blackline} which is valid for special kinds of $W$ states.}
\label{fig:ranplot1}
\end{figure}

\begin{figure}[h!]
\centering
\includegraphics[width=0.45\textwidth]{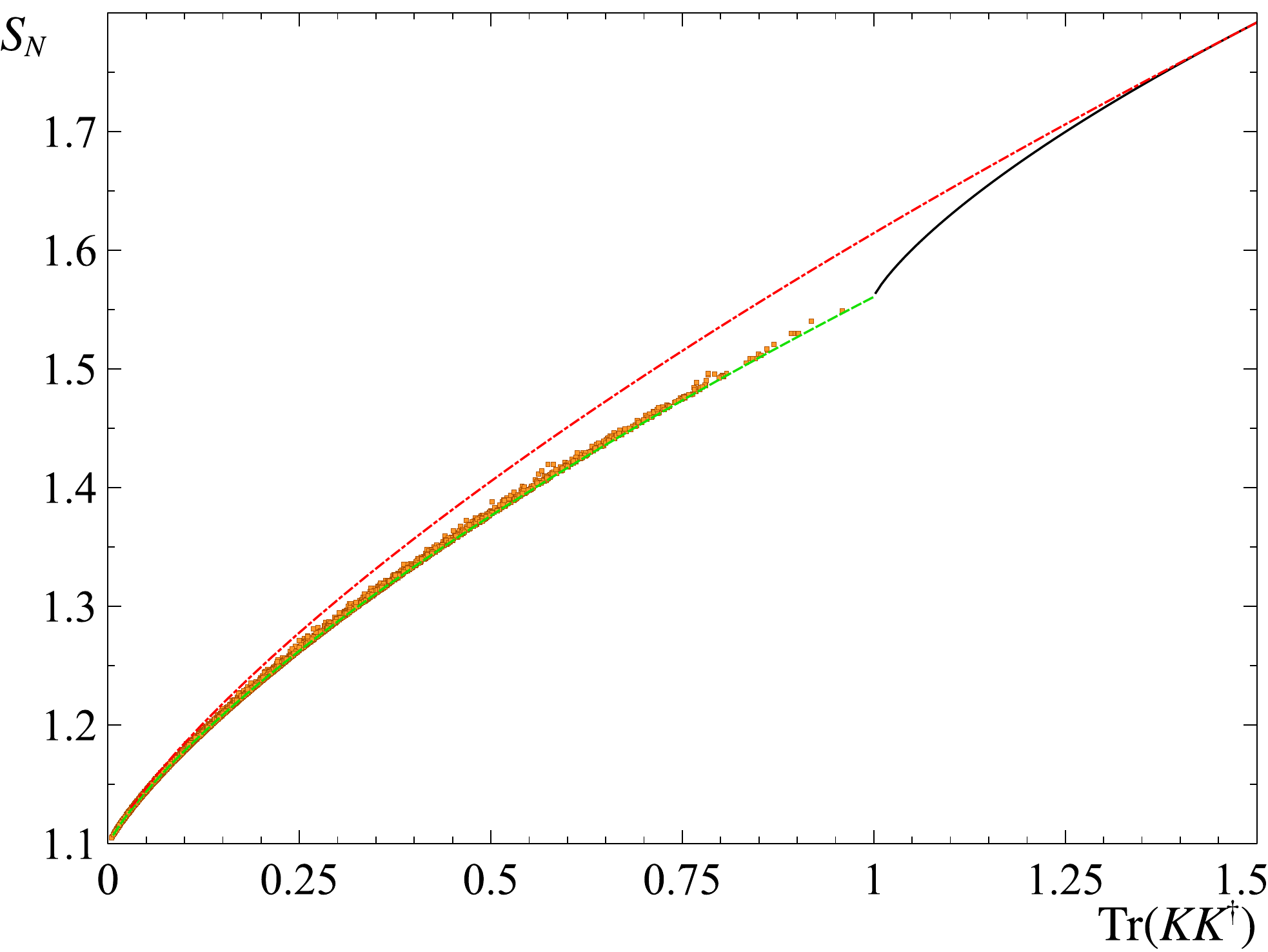}
\caption{The distribution of $\text{Tr}KK^\dagger$ and the entanglement entropy $S=-\text{Tr}\frac{\rho}{3} \log \frac{\rho}{3}$ of random generated states of three fermions with six single particle states. The orange squares are 5000 random generated states from the $W$ class. The lines are the same as in figure \ref{fig:ranplot1}.}
\label{fig:ranplot2}
\end{figure}

\begin{figure}[h!]
\centering
\includegraphics[width=0.45\textwidth]{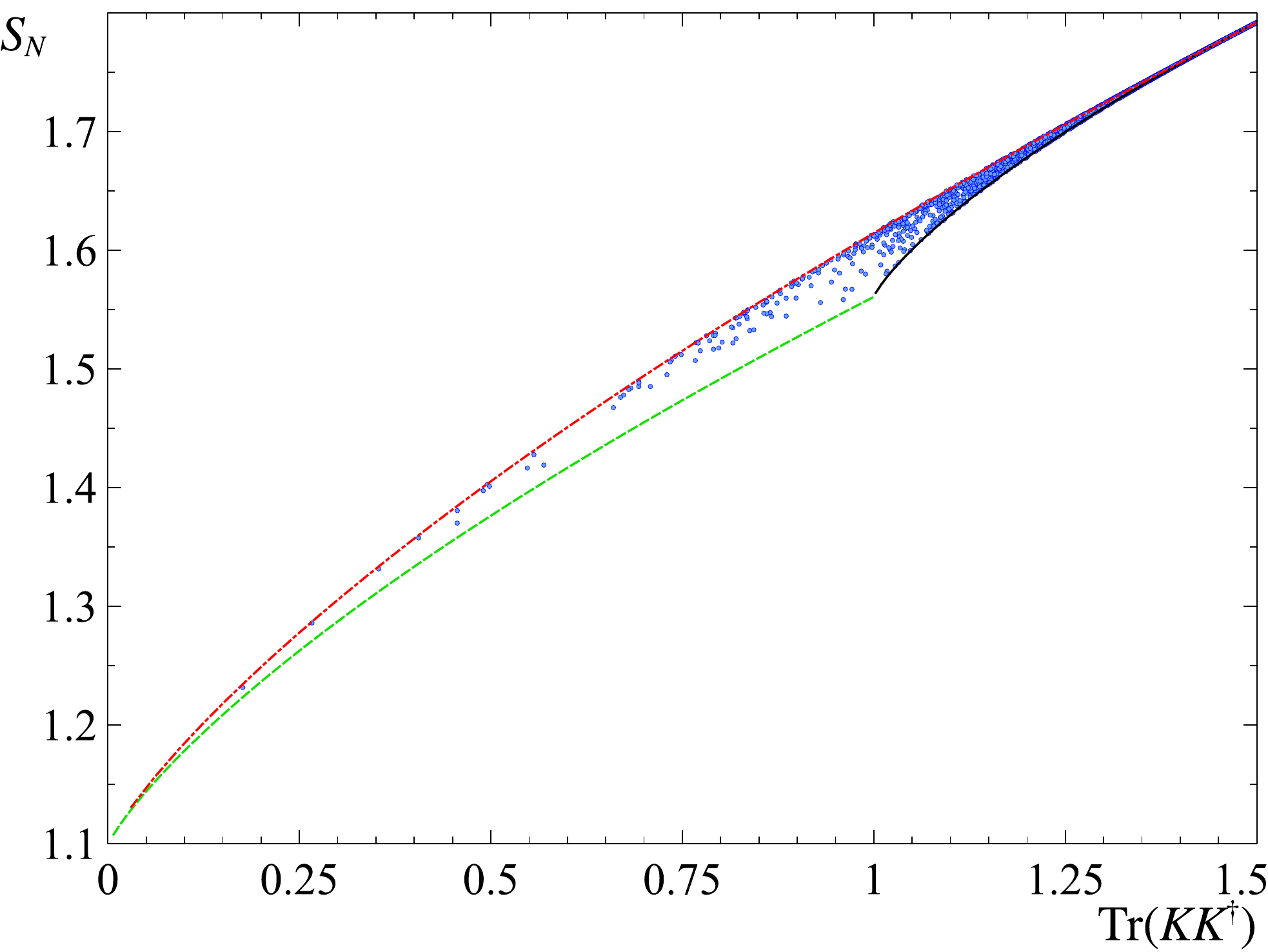}
\caption{The distribution of $\text{Tr}KK^\dagger$ and the entanglement entropy $S=-\text{Tr}\frac{\rho}{3} \log \frac{\rho}{3}$ of random generated states which are in the four parameter canonical form of e.q. \eqref{eq:canonicalform}. The blue circles are 5000 random generated states from the GHZ class. The lines are the same as in figure \ref{fig:ranplot1}.}
\label{fig:ranplot3}
\end{figure}

\begin{figure}[h!]
\centering
\includegraphics[width=0.45\textwidth]{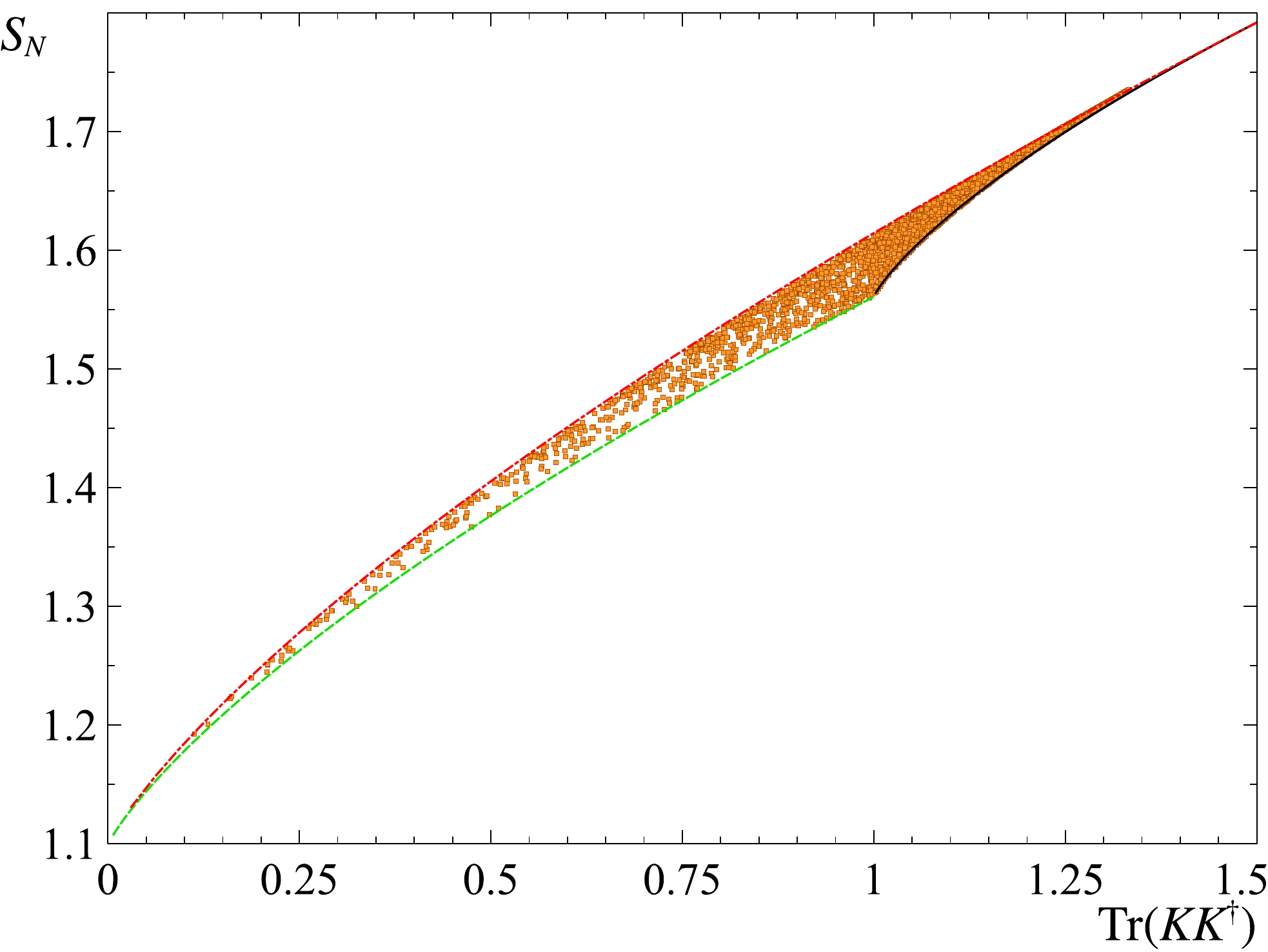}
\caption{The distribution of $\text{Tr}KK^\dagger$ and the entanglement entropy $S=-\text{Tr}\frac{\rho}{3} \log \frac{\rho}{3}$ of random generated states which are in the four parameter canonical form of e.q. \eqref{eq:canonicalform}. The orange squares are 5000 random generated states from the W class obtained by putting $\delta=0$ in \eqref{eq:canonicalform}. The lines are the same as in figure \ref{fig:ranplot1}.}
\label{fig:ranplot4}
\end{figure}

In Figure \ref{fig:ranplot1}. and \ref{fig:ranplot2}. we have plotted the quantity $\text{Tr}KK^\dagger$ and the von Neumann entropy of $\rho$ for random generated states. Figure \ref{fig:ranplot1}. contains 5000 random generated normalized states marked with blue circles. Since the GHZ orbit is dense these are definitely all GHZ states. On figure \ref{fig:ranplot2}. the orange squares are 5000 random $W$ class states obtained by acting with a random SLOCC transformation on the canonical W state. The upper bound of states with vanishing concurrence is drawn in with a dashed-dotted red line, while the entropy of biseparable states is drawn with a dashed green line. We see that the majority of GHZ states are close to the upper bound. Observe that the green line of biseparable states forms a lower bound for the entanglement entropy and the states from the $W$ class are generally close to this. In figures \ref{fig:ranplot3}. and \ref{fig:ranplot4}. we made the same plots with the difference that the states are not 
arbitarily random generated but they are in the four parameter canonical form\cite{Borland,Djok2}
\beq
\label{eq:canonicalform}
\alpha|e^{123}\rangle + \beta |e^{145}\rangle + \gamma |e^{246}\rangle + \delta |e^{356}\rangle.
\eeq
This way we can obtain a better coverage of the allowed domain. 

There is one additional curve drawn in black in figures \ref{fig:ranplot1}. and \ref{fig:ranplot2}. forming a lower bound in the region $1\leq \text{Tr}(KK^\dagger)\leq \frac{3}{2}$. This curve is the entropy of $W$ class states with four out of six eigenvalues of their one particle RDMs being $\frac{1}{2}$. Here we give a brief derivation of this curve. 

Consider equation \eqref{eq:rorotilde}. Using $K^2=\mathcal{D}(P)I$ it is easy to see that both $K$ and $K^\dagger$ commutes with $\lbrace K, K^\dagger\rbrace$. As a consequence 
\beq
[K,\rho\tilde \rho]=0.
\eeq
It follows that the eigenspaces of $\rho\tilde \rho$ are invariant subspaces of $K$. Now apart from accidental degeneracies $\rho\tilde \rho$ has two dimensional degenerate eigenspaces: the eigenvalues are $\lambda_i(1-\lambda_i)$ which are allway degenerate for pairs due to the classical Borland-Dennis result stated in e.q. \eqref{eq:BorlandDennis}. As a consequence $K$ is a block matrix of $2\times 2$ diagonal blocks in the basis where $\rho$ is diagonal. Denote these blocks with $K_\alpha$, $\alpha=1,2,3$. Using the Cayley-Hamilton theorem on $K_\alpha$ and that $K_\alpha^2=\mathcal{D}(P)I$ we see that $\left( \mathcal{D}(P)+\det K_\alpha \right)I=(\text{Tr}K_\alpha) K_\alpha$. The determinant and the trace of this equation give two equations of order two for $\det K_\alpha$ and $\text{Tr}K_\alpha$ in terms of $\mathcal{D}(P)$. Solving these equations one sees that the only solution not contradicting $\text{Tr}K=\sum_{\alpha =1}^3 \text{Tr}K_\alpha=0$ is $\det K_\alpha =-\mathcal{D}(P)$ and $\text{Tr}K_\alpha=0$. 
This shows that when $\mathcal{D}(P)=0$ the maximum possible rank of $K$ is 3. According to Table \ref{tab:1}. this is the case if we have a state in the $W$ class. Now in this case all the blocks are rank one matrices and hence can be expressed as a dyadic product of two vectors:
\beq
K_\alpha = u_\alpha v_\alpha^\dagger.
\eeq
Block diagonality ensures that vectors with different $\alpha$ index are orthogonal. We also have $u_\alpha^\dagger v_\alpha=0$ due to $K_\alpha^2=0$. This is enough to compute the anticommutator
\beq
\lbrace K,K^\dagger \rbrace = \sum_{\alpha=1}^3 |v_\alpha|^2 u_\alpha u_\alpha^\dagger + |u_\alpha|^2 v_\alpha v_\alpha^\dagger.
\eeq
Now multiply e.q. \eqref{eq:rorotilde} with the eigenvectors of $\rho \tilde \rho$ which are in this case $u_\alpha$ and $v_\alpha$ and use $\text{Tr}KK^\dagger=\sum_{\alpha=1}^3 |u_\alpha|^2|v_\alpha|^2$ to arrive at
\beq
\sum_{\beta=1}^3 |u_\beta|^2|v_\beta|^2-|u_\alpha|^2|v_\alpha|^2=4\lambda_\alpha(1-\lambda_\alpha).
\eeq
Introduce the shorthand notation $|u_\alpha|^2|v_\alpha|^2=a_\alpha$. One then has the equations
\beq
\begin{aligned}
4\lambda_1(1-\lambda_1) &= a_2+a_3\\
4\lambda_2(1-\lambda_2) &= a_1+a_3\\
4\lambda_3(1-\lambda_3) &= a_1+a_2,
\end{aligned}
\eeq
from which one can solve for the spectrum of the one particle RDM. The case when there is a pair of eigenvalues being $\frac{1}{2}$ corresponds to the equation $4\lambda_i(1-\lambda_i)=1$ between them. Now set $a_2+a_3=a_1+a_3=1$ to have four $\frac{1}{2}$ eigenvalues. Using $a_1+a_2+a_3=a_1+1=\text{Tr}KK^\dagger$ implies 
\beq
4\lambda_3(1-\lambda_3)=a_1+a_2=2 a_1=2 \text{Tr}KK^\dagger-2.
\eeq
Observe that one can only get a legitime value for $\lambda_3$ in the range $1\leq \text{Tr}KK^\dagger \leq \frac{3}{2}$. The spectrum of these kind of states as a function of $\text{Tr}KK^\dagger$ is then given as
\beq
\label{eq:blackline}
\begin{aligned}
\lambda_1 &=\lambda_6=\lambda_2 =\lambda_5=\frac{1}{2},\\
\lambda_3 &=1-\lambda_4=\frac{1}{2}\left(1+\sqrt{3-2 \text{Tr}KK^\dagger} \right).
\end{aligned}
\eeq
The entanglement entropy calculated from these eigenvalues gives the black curves of figures \ref{fig:ranplot1}. and \ref{fig:ranplot2}.

\subsection{The two particle reduced density matrix}
\label{sec:2particleRDM}

Recall that the two particle RDM has components
\beq
{\rho^{(2)}_{ij}}^{kl}=\frac{1}{2}P_{ija}\bar P^{kla}.
\eeq
A system of three fermions has the remarkable property that $\rho^{(2)}$ does not contain any additional spectral information compared to the one particle RDM $\rho$. To see this consider the following. Denote the eigenvectors of $\rho$ with $e^{(\alpha)}$:
\beq
{\rho_i}^j e^{(\alpha)}_j=\lambda^{(\alpha)}e^{(\alpha)}_i.
\eeq
It is easy to see that $E^{(\alpha)}_{ij}=P_{ijk}(\bar e^{(\alpha)})^k$, $\alpha=1,...,6$ are eigenvectors of $\rho^{(2)}$:
\beq
{\rho^{(2)}_{ij}}^{kl}E^{(\alpha)}_{kl}=P_{ija}{\bar \rho^a}_n (\bar e^{(\alpha)})^n=\lambda^{(\alpha)}P_{ija}(\bar e^{(\alpha)})^a=\lambda^{(\alpha)}E^{(\alpha)}_{ij}.
\eeq
Now the map $v^k \mapsto P_{ijk}v^k$ is a linear map from $\mathbb{C}^6$ to $\wedge^2(\mathbb{C}^6)^*$ and it can be proved that it has a SLOCC invariant rank\cite{levsar3}. A simple computation on canonical states shows that this map has full rank when $P$ is picked from the GHZ or the W classes. It follows that the eigenvectors $|E^{(\alpha)}\rangle=\frac{1}{2!}E^{(\alpha)}_{ij}{f^i}^\dagger {f^i}^\dagger|0\rangle$, $\alpha=1,...,6$ are linearly independent. Now since all $\lambda^{(\alpha)} \geq 0$ and $\text{Tr}\rho^{(2)}=3$ we conclude that the remaining nine eigenvalues are all zero and hence the spectrum of the one particle and the two particle RDM agree. It follows that all spectral based entropies also agree.

It is important to understand the physical interpretation of this result. Recall a similar well-known fact: for a bipartite system the reduced density matrices of the two subsystems have the same non-zero eigenvalues\cite{Nielsen}. This is also well-known to be true for a composite sytem of two fermions\cite{Schliemann}. In our case one can think of dividing the system of three fermions into the "bipartite" system of one fermion entangled with two fermions. Hence we have seen that in this case the non-zero eigenvalues of the reduced density matrices agree as one expects. The derivation trivialy generalizes to $N$ fermions: the non-zero eigenvalues of the $k$ particle and the $N-k$ particle RDMs agree  in general. 

Note also the important fact that we only have six non-zero eigenvalues of the two particle RDM and hence it can effectively be treated as a $6\times 6$ matrix instead of a $15\times 15$ one. We have $\binom{4}{2}=6$. This suggests that we effectively lose two single particle states and we are left with a density matrix of two fermions with four single particle states which is the fermionic system suited to describe two qubits as an embedded system.

To determine how much two-particle entanglement is present in a three particle state $|P\rangle$ one has to determine the least entangled two particle projectors on which the two particle RDM can be expanded on\cite{Wootters2}. Now we show that GHZ states with vanishing $Con(P)$ contain no two particle entanglement by showing that the natural orbitals $|E^{(\alpha)}\rangle$ with non-zero eigenvalues are separable states. Here we temporary use exterior algebra notation to simplify the formulas (for a review of this notation in the fermionic context see \cite{levsar3}). Let $V$ be a complex vector space. To show the claim recall that a $k$ fermion state $Q\in \wedge^k V$ is separable (or single Slater determinant) if and only if for every $\omega \in \wedge^{k-1}V^*$ we have\cite{Penroserindler,levsar3}
\beq
\iota_\omega Q\wedge Q=0.
\eeq
These are the so called Pl\"ucker relations\cite{Kasman}. Apply this to $Q=E^{(\alpha)}=\iota_{\bar e^{(\alpha)}} P \in \wedge^2 \mathbb{C}^6$ and $\omega=u\in (\mathbb{C}^6)^*$ to get for the condition of separability:
\beq
(\iota_u \iota_{\bar e^{(\alpha)}} P)\wedge \iota_{\bar e^{(\alpha)}} P=0,\,\,\, \forall u\in (\mathbb{C}^6)^*.
\eeq
Now using the antiderivative propierty of the interior product  we have $\iota _u(\iota_{\bar e^{(\alpha)}}P \wedge \iota_{\bar e^{(\alpha)}} P)=2(\iota_u \iota_{\bar e^{(\alpha)}} P)\wedge \iota_{\bar e^{(\alpha)}} P$ and using $\iota_{\bar e^{(\alpha)}} \circ \iota_{\bar e^{(\alpha)}}=0$ we have $\iota_{\bar e^{(\alpha)}} (\iota_{\bar e^{(\alpha)}}P \wedge P)=\iota_{\bar e^{(\alpha)}}P \wedge \iota_{\bar e^{(\alpha)}} P$. Hence the separability condition takes the form
\beq
\label{eq:sepcond}
\iota_u \iota_{\bar e^{(\alpha)}} (\iota_{\bar e^{(\alpha)}}P \wedge P)=0,\,\,\, \forall u\in (\mathbb{C}^6)^*.
\eeq
Now it is not difficult to see that $\iota_{\bar e^{(\alpha)}}P \wedge P$ is just the five-form dual to the vector ${K^i}_j ({\bar e^{(\alpha)}})^j$:
\beq
\iota_{\bar e^{(\alpha)}}P \wedge P=\frac{1}{5!}({K^i}_j(\bar e^{(\alpha)})^j) \epsilon_{iabcde} e^{abcde}.
\eeq
Using this the condition \eqref{eq:sepcond} for separability takes the form:
\beq
u^{[d}({\bar e^{(\alpha)}})^{e}{K^{a]}}_b({\bar e^{(\alpha)}})^b=0, \;\; \forall u\in (\mathbb{C}^6)^*.
\eeq
If one defines the two-form $\kappa=\frac{1}{2}({\bar e^{(\alpha)}})^{[e}{K^{a]}}_b({\bar e^{(\alpha)}})^b e_e \wedge e_a$ then this equation can be written as $u\wedge \kappa =0$ for every one-form $u=u^d e_d$. For dimensions greater than two this is equivalent with $\kappa=0$ hence
\beq
\label{eq:Keequation}
({\bar e^{(\alpha)}})^{[e}{K^{a]}}_b({\bar e^{(\alpha)}})^b=0.
\eeq
Multiply with ${e^{(\alpha)}}_e$ to get
\beq
( { e^{(\alpha)}}_e ({\bar e^{(\alpha)}})^e) {K^f}_j({\bar e^{(\alpha)}})^j=( { e^{(\alpha)}}_e {K^e}_j({\bar e^{(\alpha)}})^j)({\bar e^{(\alpha)}})^f,
\eeq
hence we see that e.q. \eqref{eq:Keequation} is equivalent with $\bar e^{(\alpha)}$ being an eigenvector of $K$. Now $K$ and $\bar \rho$ sharing all of their eigenvectors is equivalent with writing
\beq
[K,\bar \rho]=0.
\eeq
This equation is equivalent with $|E^{(\alpha)}\rangle$ being separable and hence it is a sufficient condition for $\rho^{(2)}$ to be expandable in terms of single Slater rank projections. Now we are going to show that $Con(P)=0$ implies $[K,\bar \rho]=0$. Define $C=[K,\bar \rho]$. It is sufficient to show that $\text{Tr}(CC^\dagger)=0$. Expanding $\text{Tr}(CC^\dagger)$ gives
\beq
\text{Tr}(CC^\dagger)=\text{Tr}(\bar \rho^2 \lbrace K,K^\dagger\rbrace)-2\text{Tr}(K\bar \rho K^\dagger \bar \rho).
\eeq
The term $\lbrace K,K^\dagger\rbrace$ can be expressed with $\rho$ from \eqref{eq:rorotilde}. The problematic term is $\text{Tr}(K\bar \rho K^\dagger \bar \rho)$. A long calculation shows that 
\beq
\begin{aligned}
\text{Tr}(K\bar \rho K^\dagger \bar \rho) &=\frac{1}{324}(\text{Tr}\rho)^4-\frac{1}{9}(\text{Tr}\rho^3)\text{Tr}\rho \\
&-\frac{1}{4}(\text{Tr}\rho^2)^2+\text{Tr}\rho^4 \\
&+\frac{1}{3}(\text{Tr}\rho)\text{Tr}(KK^\dagger\bar \rho)-\frac{1}{12}|\text{Tr}K^2|^2.
\end{aligned}
\eeq
We have seen that $Con(P)=0$ implies $KK^\dagger=K^\dagger K=|\mathcal{D}(P)|I$, and $\lambda^*(1-\lambda^*)=|\mathcal{D}(P)|$ for the eigenvalues of the one particle RDM (see sec. \ref{sec:vanishingconcurrence}.). Using this we write
\beq
\begin{aligned}
\label{eq:cel}
\text{Tr}(CC^\dagger) &=2\lambda^*(1-\lambda^*)\text{Tr}\rho^2-\frac{2}{324}(\text{Tr}\rho)^4 \\
&+\frac{2}{9}(\text{Tr}\rho^3)\text{Tr}\rho+\frac{2}{4}(\text{Tr}\rho^2)^2-2\text{Tr}\rho^4\\
&-\frac{2}{3}(\text{Tr}\rho)^2\lambda^*(1-\lambda^*)+\frac{2}{12}6^2(\lambda^*(1-\lambda^*))^2,
\end{aligned}
\eeq
for states with zero concurrence. Now according to Section \ref{sec:vanishingconcurrence}. $\rho$ can be written as
\beq
\rho = \left(
\begin{array}{cccccc}
 \lambda^* & & & & & \\
 & \lambda^* & & & & \\
 & & \lambda^* & & & \\
 & & & 1-\lambda^* & & \\
 & & & & 1-\lambda^* & \\
 & & & & & 1-\lambda^*
\end{array}
\right)
\eeq
when diagonalized. Substitute this into \eqref{eq:cel} to obtain the desired result
\beq
Con(P)=0 \implies \text{Tr}(CC^\dagger)=0 \implies E^{(\alpha)}\text{ are separable}.
\eeq
This result together with the fact that $Con(P)=0$ is only possible in the GHZ and and the separable classes (see section \ref{sec:vanishingconcurrence}.) suggests that $Con(P)$ measures the amount of bipartite entanglement in a three-fermion state.

\section{General considerations}
\label{sec:2}

\subsection{Fermions and qudits}
\label{sec:fermandqudits}

Before setting up the general framework for fermions and discussing the ``spin-flipped'' state it is worth saying a few words about the usefulness of studying fermionic entanglement in the task of understanding entanglement between distinguishable systems. Consider the composite system of $n$ distinguishable constituents with Hilbert spaces of dimensions $d_1$,...,$d_n$ respectively. The whole system has Hilbert space $\mathcal{H}=\mathbb{C}^{d_1}\otimes...\otimes \mathbb{C}^{d_n}$. A pure state of this system is described by the vector
\beq
|\psi\rangle = \sum_{\mu_1=1}^{d_1}...\sum_{\mu_n=1}^{d_n}\psi_{\mu_1...\mu_n}|\mu_1\rangle \otimes ... \otimes|\mu_n\rangle \in \mathcal{H}.
\eeq
Now consider a system of $n$ fermions with single particle Hilbert space $\mathcal{H}'=\mathbb{C}^{d_1}\oplus...\oplus \mathbb{C}^{d_n}$ where there is a \textit{sum} between the original Hilbert spaces instead of a product. The fermionic Hilbert space is now $\wedge^n \mathcal{H}'$. Any $|\psi\rangle \in\mathcal{H}$ can be embedded\cite{levvran3} in this space using the map
\begin{widetext}
\beq
\label{eq:generalembed}
|\psi\rangle \mapsto |P_\psi\rangle =\sum_{\mu_1=1}^{d_1}...\sum_{\mu_n=1}^{d_n}\psi_{\mu_1...\mu_n} {f^{\mu_1}}^\dagger {f^{d_1+\mu_2}}^\dagger...{f^{d_1+...+d_{n-1}+\mu_n}}^\dagger|0\rangle.
\eeq
\end{widetext}
This embedding has the nice propierty that it also embeds the SLOCC group of the distinguishable system into the fermionic one in a nice way. The SLOCC group of the distinguishable system is $GL(d_1,\mathbb{C})\otimes ...\otimes GL(d_n,\mathbb{C})$ acting locally as
\beq
\begin{aligned}
 \psi_{\mu_1...\mu_n} &\mapsto {(g_1)^{\nu_1}}_{\mu_1}...{(g_1)^{\nu_n}}_{\mu_n}  \psi_{\nu_1...\nu_n},\\
 g_1\otimes ...\otimes g_n &\in GL(d_1,\mathbb{C})\otimes ...\otimes GL(d_n,\mathbb{C}).
\end{aligned}
\eeq
These transformations are implemented via the fermionic SLOCC transformations
\beq
g=
\left(
\begin{array}{ccc}
g_1 & & \\
 & \ddots & \\
 & & g_n 
\end{array}
\right) \in GL(\mathcal{H}')
\eeq
This shows that states of the embedded system on the same $GL(d_1,\mathbb{C})\otimes ...\otimes GL(d_n,\mathbb{C})$ orbit are on the same $GL(\mathcal{H}')$ orbit when considered as fermionic states. Hence the entanglement classes of the fermionic system give a coarse graining of the classes of the distinguishable system. As we have seen this coarse graining is a one to one correspondence in the case of three qubits and three fermions with six single particle states and in the case when only bipartite entanglement is considered but it works remarkably well in the case of other multipartite systems as well, for example four qubits embedded into four fermions with eight single particle states\cite{DjokPRA}, or three qudits embedded into three fermions with nine single particle states\cite{levsar3}.

Note that the embedding \eqref{eq:generalembed} has a nice physical interpretation. Consider $n$ nodes where fermions can be localized and on the $k$th node a fermion can have $d_k$ internal states. These nodes can be energy levels of atoms or nodes in a lattice or anything alike. The subspace of the $n$ fermion Hilbert space $\mathcal{H}'$ defined by \eqref{eq:generalembed} is just the single occupancy subspace of this node interpretation: where we allow only states where on each node there is exactly one fermion. If we prescribe this condition it is clear that the fermions suddenly become distinguishable and the resulting Hilbert space is just $\mathcal{H}$. 

Despite how natural the above physical interpretation seems, the embedding \eqref{eq:generalembed} is by no ways canonical when the whole Fock space of fermions is considered. For example one can use double occupancy states instead of single occupancy ones and end up with a good embedding\cite{levsar2} of the distinguishable SLOCC group into the extended fermionic SLOCC group, which we review in the next subsection. 

\subsection{The extended SLOCC group}
\label{sec:extendedSLOCC}

In this subsection we briefly review the concept of the extended fermionic SLOCC group which was recently introduced by us\cite{levsar2}. For an extensive treatment we refer to this work. The mathematics used to define this concept is known long ago and called the classification of spinors\cite{Chevalley,Igusa}. The physical interpretation is that we define states related by a Bogoliubov transformation to be equally entangled. This way we classify fermionic Fock spaces without a reference to a specific vacuum which is only special if a Hamiltonian is specified. 

Let $\mathcal{H}$ be a $d$ dimensional one particle Hilbert space. One constructs the Fock space of fermions as
\beq
\mathcal{F}=\mathbb{C}\oplus \mathcal{H} \oplus \wedge^2 \mathcal{H} \oplus ... \oplus \wedge^d \mathcal{H}.
\eeq
It is clear that $\dim \mathcal{F}=2^d$. The term $\mathbb{C}$ in the direct sum is spanned by the so called vacuum and is denoted by $|0\rangle$. Let $e^i$ be a basis of $\mathcal{H}$ and $e_i$ the dual basis of $\mathcal{H}^*$ satisfying $e_i(e^j)={\delta_i}^j$. Note that we have temporary dropped the bracket notation for the vectors in the one particle Hilbert space and its dual because we would like to introduce the Hermitian inner product of $\mathcal{H}$ later, to emphasize what structures are pre-existing on $\mathcal{F}$ regardless of the choice of a Hermitian inner product. Also, in a somewhat unorthodox way, we use upper indices for basis vectors of $\mathcal{H}$ and lower ones for its dual. This is done to conform with the notation used in the study of exterior algebras. The creation operators assoiciated with $e^i$ are denoted with $p^i$ and the annihilation operators associated with $e_i$ are denoted with $f_i$. We temporary use $p$ instead of $f^\dagger$ since we reserve $\dagger$ for the 
adjoint 
with respect to the Hermitian inner product. These satisfy $\lbrace p^i,p^j\rbrace=\lbrace f_i,f_j\rbrace=0$ and
\beq
\lbrace p^i,f_j\rbrace ={\delta_i}^j I.
\eeq
Here $\lbrace.,.\rbrace$ denotes the usual anticommutator. Denote the vector space of creation operators spanned by $p^i$ with $W$. The dual vector space $W^*$ is spanned by the annihilation operators and the dual action is defined through
\beq
\lbrace p,f\rbrace = f(p)I, \;\; p\in W,\;f\in W^*.
\eeq
Now consider the vector space $W\oplus W^*$ spanned by creation and annihilation operators together. Since the anticommutator of two such object is always proportional to the identity, $W\oplus W^*$ is naturally endowed with an inner product $(.,.)$ defined via
\beq
\label{eq:Cliff}
\lbrace X,Y\rbrace = 2(X,Y)I.
\eeq
If we pick coordinates $X=x_i p^i+u^jf_j$ and $Y=y_ip^i+v^j f_j$ we can write $(X,Y)=\frac{1}{2}(x_i v^i + u^i y_i)$. Notice that the relation \eqref{eq:Cliff} is the \textit{defining} relation of the Clifford algebra of an inner product space. Hence the creation and annihilation operators generate the Clifford algebra $Cliff(W\oplus W^*)$. Note that the general result\cite{Chevalley} $Cliff(W\oplus W^*)\cong End(\wedge^\bullet W) \cong End(\mathcal{F})$ implies that every endomorphism of the Fock space can be obtained as an element of the Clifford algebra generateted by creation and annihilation operators. 

The orthogonal group of the inner product $(.,.)$ is $SO(W\oplus W^*)$:
\beq
(\mathcal{O}(X),\mathcal{O}(Y))=(X,Y), \;\; \forall \mathcal{O} \in SO(W\oplus W^*).
\eeq
By construction the action of this group leaves the anticommutator invariant: these are the Bogoliubov transformations of the space $W\oplus W^*$ of creation and annihilation operators. In order to classify states in $\mathcal{F}$ we would like to obtain an action of this group on $\mathcal{F}$. A natural way to do this is to define the operator $O$ acting on $\mathcal{F}$ via the relation
\beq
\label{eq:doublecover}
OXO^{-1}=\mathcal{O}(X), \;\; X\in W\oplus W^*, \; \mathcal{O}\in SO(W\oplus W^*).
\eeq
However this definition is not unique: $O$ and $-O$ both satisfy this for the same $\mathcal{O}$. On the other hand if we allow both we end up with the double cover $Spin(W\oplus W^*)$ of $SO(W\oplus W^*)$ acting on the fermionic Fock space $\mathcal{F}$. To get a feeling of this action it is instructive to work out the infinitesimal version, where the correspondence between $Spin(W\oplus W^*)$ and $SO(W\oplus W^*)$ is one to one since they share the Lie algebra $\mathfrak{so}(W\oplus W^*)$. It is easy to see that the infinitesimal version of \eqref{eq:doublecover} is
\beq
\label{eq:doublecoverinf}
[T,X]=\mathcal{T}(X),
\eeq
where $\mathcal{T}$ satisfies $(\mathcal{T} X,Y)+(X,\mathcal{T}Y)=0$. This latter condition can be parametrized as
\beq
\label{eq:Tdefinieing}
\begin{aligned}
\mathcal{T}
\left(
\begin{array}{c}
p^i \\
f_j
\end{array}
\right)
&=
\left(
\begin{array}{cc}
A^i_{\;k} & \beta^{il} \\
B_{jk} & -A^l_{\;j}
\end{array}
\right)
\left(
\begin{array}{c}
p^k \\
f_l
\end{array}
\right), \\ B_{ij} &=-B_{ji}, \; \beta^{ij}=-\beta^{ji}.
\end{aligned}
\eeq
It is easy to see that the operator implementing \eqref{eq:doublecoverinf} is 
\beq
\label{eq:generators}
T=\frac{1}{2}{A^i}_j[p^j,f_i]-\frac{1}{2}B_{ij}p^ip^j-\frac{1}{2}\beta^{ij}f_if_j.
\eeq
We parametrize elements from the identity component of $Spin(W\oplus W^*)$ as $O=e^T$. The action on a state $|\psi\rangle \in \mathcal{F}$ is
\beq
|\psi\rangle \mapsto e^T|\psi\rangle.
\eeq
We define the extended SLOCC group to be $Spin(W\oplus W^*)$ and the entanglement classes of $\mathcal{F}$ to be the orbits of this group with respect to the above action on states. It is a natural extension for the following main reasons.
\begin{itemize}
 \item The particle number conserving subgroup is obtained by setting $B_{ij}=\beta^{ij}=0$. This acts on creation operators as $O p^i O^{-1}={g^i}_jp^j$ where the matrix ${g^i}_j$ is the exponential of the otherwise unconstrained matrix ${A^i}_j$ and hence $g\in GL(W)\cong GL(\mathcal{H})$. The action on Slater determinants is
 \beq
 p^{i_1}...p^{i_m}|0\rangle \mapsto (\det g)^{-1/2}g^{i_1}_{\;k_1}p^{k_1}...g^{i_m}_{\;k_m}p^{k_m}|0\rangle,
\eeq
which is appart from the factor $(\det g)^{-1/2}$ is the usual action of the fermionic SLOCC group.

\item The states $|\psi\rangle\in \mathcal{F}$ for which the annihilator subspace $E_\psi =\lbrace X\in W\oplus W^*|\;X|\psi\rangle =0\rbrace$ is of maximal dimension are called pure spinors\cite{Chevalley}. These are considered to be the least entangled states in this classification. Note that every $X\in E_\psi$ is nilpotent since $0=X^2|\psi\rangle =(X,X)|\psi\rangle$. It also follows that $E_\psi$ contains isotropic vectors and hence its maximal dimension is $d$. It follows that every pure spinor $|\psi\rangle$ can be regarded as a vacuum state with the operators spanning $E_\psi$ regarded as its annihilation operators. Conversely, every possible vacuum has a $d$ dimensional annihilator subspace spanned by its annihilation operators and hence is a pure spinor. It can be shown that every pure spinor can be obtained as a $B$-transform (a spin transform with $A=\beta=0$) of a Slater determinant. Pure spinors always form a single orbit under $Spin(W\oplus W^*)$.

\item The action of $Spin(W\oplus W^*)$ on $\mathcal{F}$ is reducible with irreducible subspaces $\mathcal{F}=\mathcal{F}^+\oplus \mathcal{F}^-$ where $\mathcal{F}^+$ is the even and $\mathcal{F}^-$ is the odd particle subspace. Hence this classification naturally prevents mixing between fermionic and bosonic multiparticle states.
\end{itemize}

When we restore the Hermitian inner product $\langle .|.\rangle$ we can restrict ourselves to $Spin(W\oplus W^*)$ transformations which are unitary with respect to this inner product. This defines the extended LOCC group. It can be shown\cite{levsar2} that it is the compact real form of $Spin(W\oplus W^*)$, which is $Spin(2d)$.

\subsection{The invariant bilinear product}
\label{sec:bilinprod}

There is a canonical $Spin(W\oplus W^*)$ invariant bilinear product\cite{Chevalley} on the Fock space $\mathcal{F}$ which is defined regardless of the existence of a Hermitian inner product. To construct this, first we need the transpose $t$ of a Clifford algebra element. It is defined to be 
\beq
\label{eq:transpose}
(X_1 X_2 ...X_k)^t=X_k...X_2 X_1,
\eeq
for elements which are products of vectors $X_i\in W\oplus W^*$ and extended linearly on $Cliff(W\oplus W^*)$. This naturaly defines the transpose of elements of the Fock space. For a Slater determinant we define
\beq
(p^{i_1}...p^{i_k}|0\rangle)^t = (p^{i_1}...p^{i_k})^t|0\rangle\equiv (-1)^{\frac{k(k-1)}{2}}p^{i_1}...p^{i_k}|0\rangle,
\eeq
and extend this to $\mathcal{F}$ linearly. The invariant bilinear product $(\phi,\psi)$ of states $|\phi\rangle$, $|\psi\rangle \in \mathcal{F}$ is defined as
\beq
\label{eq:Mukai}
(\phi,\psi)=(|\phi\rangle^t\wedge |\psi\rangle)_{\text{top}},
\eeq
where the subscript top indicates the coefficient multiplying the state $|top\rangle = p^1...p^d|0\rangle$, the top state with all single particle states filled. Note that by an abuse of notation we used $(.,.)$ for this product as well as the one living on $W\oplus W^*$ defined in e.q. \eqref{eq:Cliff}. It should always be clear from the context which one we are referring to. Note that the symmetry of the pairing depends on the dimension of the one particle Hilbert space:
\beq
\label{eq:symMukai}
(\phi,\psi)=(-1)^{\frac{d(d-1)}{2}} (\psi,\phi).
\eeq
To see the invariance of this product it is useful to write it in a different form. Define $\Omega = f_1...f_d$. It is easy to see that every state is annihilated by $\Omega$ except the subspace of $|top\rangle$:
\beq
\Omega |top\rangle = (-1)^{\frac{d(d-1)}{2}}|0\rangle.
\eeq
We can write \eqref{eq:Mukai} equivalently as
\beq
(\phi,\psi)|0\rangle = (-1)^{\frac{d(d-1)}{2}}\Omega |\phi\rangle^t\wedge |\psi\rangle
\eeq
Now let $\Psi \in Cliff(W\oplus W^*)$ be an operator which creates $|\psi\rangle$ from the vacuum i.e. $|\psi\rangle =\Psi |0\rangle$. Of course $\Psi$ is not uniquely determined but this is not a requirement now. Chose a $\Phi$ the same way for $|\phi\rangle$. Then we have 
\beq
\label{eq:Mukai2}
(\phi,\psi)|0\rangle=(-1)^{\frac{d(d-1)}{2}}\Omega \Phi^t \Psi |0\rangle.
\eeq
With this form one can easily see that for any $A\in Cliff(W\oplus W^*)$ one has 
\beq
(\phi,A\psi)=(A^t\phi,\psi),
\eeq
hence the name transpose is justified. Now for the generators of the spin group defined in e.q. \eqref{eq:generators} we have $T^t=-T$. This implies
\beq
(\phi,T\psi)+(T\phi,\psi)=0,
\eeq
which proves the invariance. With the use of this invariant bilinear product one can associate covariants to a state transforming as tensors under $SO(W\oplus W^*)$. For example the quantities $(\psi, p^i \psi)$, $(\psi, f_j\psi)$ together transform as a vector. The covariants can be used to find the orbit structure of the Fock space and to construct continous invariants which then serve as entanglement measures (see \cite{levsar2} for details). 

\subsection{The ``spin-flipped'' dual state}
\label{sec:spinflipp2}

 Introduce a Hermitian inner product $\langle.|.\rangle$ on $\mathcal{H}$ and extend it to $\mathcal{F}$ in the usual way. This introduces a complex structure on the space $W\oplus W^*$ of creation and annihilation operators: for an orthonormal basis $\langle e^i |e^j\rangle ={\delta^i}_j$ one has $p^i=(f_i)^\dagger$: the creation operators are the adjoints of the annihilation operators. In a more formal way for every state $v_ie^i \in \mathcal{H}$ one can associate a creation operator $p_v=v_ip^i$ and for every dual state $u^i e_i \in \mathcal{H}^*$ one can associate an annihilation operator $f_u=u^if_i$. The inner product introduces an antilinear map $A:\mathcal{H}\rightarrow \mathcal{H}^*$ defined as $A(v)(w)=\langle v|w\rangle$. Then the adjoint defined from this inner product satisfies $(f_{A(v)})^\dagger = p_v$.

The fermionic generalization of the ``spin-flipped'' state is just a dual state relating the Hermitian inner product with the invariant bilinear pairing of e.q. \eqref{eq:Mukai}. We define the antilinear automorphism $\chi:\mathcal{F} \rightarrow \mathcal{F}$ of the Fock space as
\beq
\label{eq:chidef}
\langle \chi \phi |\psi \rangle = ( \phi,\psi), \;\; \forall \phi,\psi \in \mathcal{F}.
\eeq
The ``spin-flipped'' dual of $|\phi\rangle$ is then $|\chi \phi \rangle$. Using the form \eqref{eq:Mukai2} with the assumption that we chose a normalized vacuum we can write this as
\beq
\langle 0|(\chi\Phi)^\dagger \Psi|0\rangle= (-1)^{\frac{d(d-1)}{2}}\langle 0| \Omega \Phi^t \Psi |0\rangle,
\eeq
from where we can read off the action of $\chi$ on an arbitary state:
\beq
\label{eq:chi0}
|\chi \phi\rangle= (-1)^{\frac{d(d-1)}{2}} (\Omega \Phi^t)^\dagger |0\rangle.
\eeq
Now it is easy to see that $\Omega^\dagger = f_d^\dagger...f_1^\dagger = (-1)^{\frac{d(d-1)}{2}}f_1^\dagger...f_d^\dagger$ and hence $(-1)^{\frac{d(d-1)}{2}}\Omega^\dagger |0\rangle = |top\rangle$. One is left with
\beq
\label{eq:chi}
|\chi \phi\rangle = (\Phi^t)^\dagger |top\rangle.
\eeq
The adjoint conjugates every amplitude in $\Phi$ and changes creation operators to annihilation ones with reversing the order. The transpose restores the original order i.e. for $\Phi=\phi^{(0)}+\phi^{(1)}_i {f^i}^\dagger+\frac{1}{2}\phi^{(2)}_{ij}{f^i}^\dagger {f^j}^\dagger +...$ one has $(\Phi^t)^\dagger=\bar \phi^{(0)}+(\bar \phi^{(1)})^i f_i+\frac{1}{2}(\bar \phi^{(2)})^{ij} f_i f_j +...$. This shows that $|\chi\phi\rangle$ is nothing but an interesting particle-hole dual of $|\phi\rangle$ with an additional complex conjugation: the complex conjugate state is annihilated out of the fully filled state. This picture is reassured if one calculates the action on Slater determinant states:
\begin{widetext}
\beq
\chi({f^{i_1}}^\dagger...{f^{i_k}}^\dagger|0\rangle) = (-1)^{\frac{k(k-1)}{2}} \frac{1}{(d-k)!} \epsilon_{i_1...i_kj_{k+1}...j_d}{f^{j_{k+1}}}^\dagger...{f^{j_d}}^\dagger|0\rangle.
\eeq
\end{widetext}
From e.q. \eqref{eq:chi0} it is easy to see that $\chi^2=(-1)^{\frac{d(d-1)}{2}}$. A mathematicaly interesting consequence is that either $\chi$ or $i\chi$ is an antilinear involution and hence a complex structure. Since $\chi$ is uniquely defined by the Hermitian inner product it follows that fixing a Hermitian inner product on $\mathcal{H}$ is equivalent with fixing a complex structure on $\mathcal{F}$.

Since unitary invariants are calculated from the Hermitian inner product, while SLOCC invariants are calculated from the invariant bilinear pairing one sees that the role of the ``spin-flip'' dual is to allow one the calculation of SLOCC invariants from the Hermitian inner product. It is trivial that the dual state $|\tilde P \rangle$ of three fermions with six single particle states defined in \eqref{eq:Pdual} is just the image of $|P\rangle$ under $\chi$. The reduced density matrix of $| \tilde P \rangle$ gave rise to the ``spin-flipped'' density matrices of three qubits under the embedding \eqref{eq:3qubitembedding}. The elements of the one particle reduced density matrix are given by ${\rho^i}_j=\langle P|{f^i}^\dagger f_j P\rangle$ and one can check that the elements of the matrix $K$ defined in \eqref{eq:Kmatrix} are just ${K^i}_j = (P,{f^i}^\dagger f_j P)$. 

The relation \eqref{eq:dualhole} suggesting the particle-hole picture for the ``spin-flipped'' RDM can now be proved generaly. Using that $\chi^{-1}=(-1)^{\frac{d(d-1)}{2}}\chi$ and e.q. \eqref{eq:symMukai} it is easy to see that $\chi$ is antiunitary: $\langle \chi \phi|\chi \psi\rangle = \langle \psi |\phi\rangle$. Using this and \eqref{eq:chi} the following result is straightforward
\beq
\langle \chi \psi | A \chi \psi \rangle = \langle \psi | A^t \psi \rangle.
\eeq
Since $({f^i}^\dagger f_j)^t= f_j {f^i}^\dagger={\delta^i}_j-{f^i}^\dagger f_j$ we indeed have
\beq
\tilde { \rho^i}_j=\langle \chi \psi | {f^i}^\dagger f_j \chi \psi \rangle ={\delta^i}_j-\langle \psi | {f^i}^\dagger f_j \psi \rangle= {\delta^i}_j-{\rho^i}_j,
\eeq
for normalized states.

To conclude this section we derive the general relations between reduced density matrix elements and SLOCC covariants. The relation of e.q. \eqref{eq:rorotilde} is a special case of these relations. To derive these relations first note that there are two type of projections defined from the two inner products. The first is the usual one defining density matrices from pure states:
\beq
\begin{aligned}
P_\psi &: \mathcal{F} \rightarrow \mathcal{F}, \\
& |\phi\rangle \mapsto \langle \psi |\phi\rangle |\psi\rangle.
\end{aligned}
\eeq
We usually write $P_\psi =|\psi\rangle \langle \psi|$. The other one is defined from the invariant bilinear product:
\beq
\begin{aligned}
{P'}_\psi &: \mathcal{F} \rightarrow \mathcal{F}, \\
& |\phi\rangle \mapsto ( \psi ,\phi) |\psi\rangle.
\end{aligned}
\eeq
From the definition \eqref{eq:chidef} one sees that this can be written as ${P'}_\psi=|\psi\rangle \langle \chi \psi|$. On the other hand from \eqref{eq:Mukai2} we have ${P'}_\psi = (-1)^{\frac{d(d-1)}{2}} \Psi \Omega \Psi^t$. We would now like an expansion of $P_\psi$ and ${P'}_\psi$ in terms of density matrix elements and SLOCC covariants respectively. To obtain this we employ that since $Cliff(W\oplus W^*)\cong End(\mathcal{F})$ the trace of an arbitary Clifford algebra element is well-defined. Let $\lbrace |\vartheta_i\rangle\rbrace_{i=1}^{2^d}$ be a basis of $\mathcal{F}$ and $|{\vartheta^*}_i\rangle$ be the dual basis with respect to the product $\langle .|.\rangle$ while $|{\vartheta^{**}}_i\rangle$ be the dual basis with respect to the product $(.,.)$. Obviously $|{\vartheta^*}_i\rangle=|\chi {\vartheta^{**}}_i\rangle$. We have
\beq
\begin{aligned}
\text{tr} A &=\sum_i \langle {\vartheta^*}_i | A \vartheta_i \rangle = \sum_i ({\vartheta^{**}}_i,A \vartheta_i), \\
A &\in Cliff(W\oplus W^*).
\end{aligned}
\eeq
It is obvious that 
\beq
\label{eq:Trkif1}
\text{Tr}(A P_\psi)=\langle \psi |A\psi\rangle,
\eeq
and
\beq
\label{eq:Trkif2}
\text{Tr}(A {P'}_\psi)=( \psi ,A\psi).
\eeq
Now choose a basis $\lbrace \theta_i \rbrace_{i=1}^{2^{2d}}$ of $Cliff(W\oplus W^*)$ and denote its trace-dual with $\theta^i$ i.e. $\text{tr}\theta^i \theta_j={\delta^i}_j$. A self-dual basis for example can be obtained from a gamma matrix basis of $W\oplus W^*$ as $\lbrace \frac{1}{k!}\gamma_{[i_1}...\gamma_{i_k]} | \lbrace i_1,...,i_k\rbrace \subseteq \lbrace 1,...2d\rbrace ,\;0\leq k\leq 2d \rbrace$. A gamma matrix basis of $W\oplus W^*$ is a basis which is orthonormal with respect to the inner product \eqref{eq:Cliff} of $W\oplus W^*$, i.e. $\lbrace \gamma_i,\gamma_j\rbrace=2\delta_{ij}$. Using such a basis and e.q. \eqref{eq:Trkif1} and \eqref{eq:Trkif2} with $A=\theta_i$ it is straightforward to get the expansions
\beq
\begin{aligned}
P_\psi &=|\psi\rangle \langle\psi|= \sum_i \langle \psi|\theta_i\psi \rangle \theta^i, \\
{P'}_\psi &=|\psi\rangle \langle \chi \psi|= \sum_i ( \psi,\theta_i\psi ) \theta^i. 
\end{aligned}
\eeq
Using these we can easily derive our final results
\beq
\begin{aligned}
(\psi,A\psi)\overline{(\psi,B\psi)} &=\langle B \psi|\chi \psi\rangle \langle \chi \psi |A\psi\rangle \\
&=\sum_i \langle \psi |B^\dagger \theta^i A \psi \rangle \langle \chi \psi|\theta_i \chi \psi\rangle,
\end{aligned}
\eeq
and
\beq
\langle \psi |A\psi\rangle \langle \chi \psi|B \chi \psi \rangle = \sum_i (\psi, \theta_i \psi) \overline{(\psi, (A \theta^i B)^\dagger \psi)},
\eeq
valid for all $A,B\in Cliff(W\oplus W^*)$. Physically speaking the first relation expresses that a SLOCC covariant multiplied by its conjugate can be expanded with the use of reduced density matrix elements of the state and its ``spin-flipped'' dual, while the second relation is basically the inverse of the first. Mathematically speaking these equations relate spinor bilinears between the spinor $|\psi\rangle$ and its dual $|\chi\psi\rangle$ with the bilinears of these spinors with themselves. These kinds of relations between two spinors are called Fierz identities\cite{Fierz,Fierz2} in the theory of spinors.

\section{Summary}
\label{sec:concl}

In the first half of this paper we have derived an inequality for the absolute value of the quartic invariant of three fermions with six single particle states which reduces to the inequality of Coffman, Kundu and Wootters when three qubit-like states are considered. Motivated by this we have defined a concurrence for pure three fermion states as $Con(P)=(3-\text{Tr}\rho^2)-6|\mathcal{D}(P)|\geq 0$. We have shown that the vanishing of this quantity is only possible in the GHZ and the separable classes and that it implies that the two particle RDM is a mixture of separable states. We gave bounds on the entanglement entropy in terms of the entropy $3-\text{Tr}\rho^2\geq 0$ and showed that they are almost interchangeable. Hence we argued that the equation
\beq
3-\text{Tr}\rho^2=6|\mathcal{D}(P)|+Con(P)\leq \frac{3}{2}
\eeq
expresses that the amount of entanglement of a fermion with the rest of the system is the sum of the amount of its bipartite and tripartite entanglement.

In the second half of the paper we have related SLOCC covariants with reduced density matrix elements (or local unitary covariants) for general fermionic systems. The bridge between these quantities is a conjugate particle-hole dual state. This dual state is the fermionic generalization of the "spin-flipped" dual for qubits and hence may give a nice, more fundamental physical interpretation for the latter. When one regards fermionic states as spinors the relations between SLOCC covariants and reduced density matrix elements are Fierz identities between the state and its dual. Since in the first half of the paper we have seen that the CKW inequality originates from such a relation we suspect that every monogamy-like inequality can be tracked back to have roots in a Fierz identity. Perhaps these findings give another good argument for studying entanglement between fermions: the well-known tools used to study entanglement find a unified origin in the beautiful mathematical theory of spinors. 

There are still many open questions and possible work to be done in understanding multipartite fermionic entanglement. The SLOCC covariants classifying three fermions with seven, eight and nine single particle states\cite{levsar3} and four fermions with eight single particle states\cite{DjokPRA} are well known. These play the role of the matrix $K$ and one can derive similiar inequalities as the one in this paper. However, finding physical interpretation of the resulting quantities probably requires a more involved analysis. The system of three fermions with nine single particle states and four fermions with eight single particle states are particulary interesting since they contain the distinguishable system of three qutrits and four qubits respectively. Inequalities for these systems would provide inequalities for the corresponding distinguishable systems too. There is also much work to be done regarding the spinor formalism of fermionic entanglement. For example the particle-hole dual state can be 
generalized to mixed states: a 
mixed state $\rho\in Cliff(W\oplus W^*)$ has a canonical dual defined with the transpose of e.q. \eqref{eq:transpose} as $\tilde \rho=\rho^t$. It is not difficult to see that while neither $\rho$ nor $\tilde \rho$ transforms covariantly under general SLOCC transformations (just under LOCC) the quantity $\rho \tilde \rho$ does so. This opens the possibility of generating SLOCC covariants and invariants for mixed states and might allow one to go beyond the usual convex roof method when classifying mixed state entanglement.


\begin{thebibliography}{}

\bibitem{Nielsen} M. A. Nielsen and I. L. Chuang: Quantum Information and Quantum computation, Cambridge University Press, 2000
\bibitem{Ryu} S. Ryu, T. Takayanagi, Phys. Rev. Lett. {\bf 96}, 181602, 2006.
\bibitem{Coffman} V. Coffman, J. Kundu, W. K. Wootters, Phys. Rev. A{\bf 61}, 052306, 2000.
\bibitem{Dur} W. D\"ur, G. Vidal, J. I. Cirac, Phys. Rev. A{\bf 62} (2000) 062314.
\bibitem{Bennett} C. H. Bennett, S. Popescu, D. Rohrlich, J. A. Smolin and A. V. Thapliyal, Phys. Rev. A{\bf63}, 012307 (2000).
\bibitem{Zanardi} P. Zanardi, Phys. Rev. A{\bf 65}, 042101 (2002).
\bibitem{Banuls} M. C. Banuls, J. I. Cirac and M. M. Wolf, Phys. Rev. A {\bf 76}, 022311 (2007).
\bibitem{Heaney} L. Heaney and V. Vedral, Phys. Rev. Lett. 103, 200502 (2009).
\bibitem{Coleman} A. J. Coleman, Rev. Mod. Phys. {\bf 35} 668-686, 1963.
\bibitem{Borland} R. E. Borland and K. Dennis, Journal of Physics B{\bf 5}, 7-15, 1972.
\bibitem{Kly1} A. A. Klyachko, Journal of Physics: Conference Series {\bf 36}, 2006.
\bibitem{levvran1} P. L\'evay and P. Vrana, Phys. Rev. A{\bf 78} (2008), 022329.
\bibitem{Kimura} T. Kimura, {\it Introduction to Prehomogeneous Vector spaces}
Translations of Mathematical Monographs. Volume 215, American
Mathematical Society, (2003).
\bibitem{Satokimura} M. Sato and T. Kimura, Nagoya Math. J. {\bf 65} 1-155 (1977).
\bibitem{levsar2} G. S\'arosi and P. L\'evay, J. Phys. A: Math. Theor. {\bf 47} 115304 (2014).
\bibitem{levsar3} G. S\'arosi and P. L\'evay, Phys. Rev. A {\bf 89} 042310 (2014).
\bibitem{Chevalley} C. Chevalley, {\it The algebraic Theory of Spinors}, Columbia University Press, 1954.
\bibitem{Cayley} A. Cayley, Camb. Math. 4, pp. 193-209, (1845).
\bibitem{Penroserindler} R. Penrose and W. Rindler, {\it Spinors and Space-Time Vol 1.}, Cambridge Monographs on Mathematical Physics, Cambridge University Press 1984.
\bibitem{Kasman} A. Kasman, T. Shiota, K. Pedings and A. Reiszl, The Proceedings
of the American Mathematical Society {\bf 136} 77-87 (2008).
\bibitem{DjokPRA}
L. Chen, D. Z. Djokovic, M. Grassl, and B Zeng, Phys. Rev. A{\bf 88} 052309 (2013).
\bibitem{Djok2} L. Chen, D. Z. Djokovic, M. Grassl and B. Zeng, J. Math. Phys. {\bf 55} 082203 (2014). 

\bibitem{Kempe} J. Kempe, Phys. Rev. A {\bf 60} 910 (1999).
\bibitem{Hill} S. Hill and W. K. Wootters, Phys. Rev. Lett. {\bf 78} 5022 (1997).
\bibitem{Wootters} W. K. Wootters, Phys. Rev. Lett. {\bf 80} 2245 (1998).
\bibitem{levvran3} P. Vrana and P. Levay, J. Phys. A: Math. Theor. {\bf 42}, 285303 (2009).
\bibitem{Igusa} J.I. Igusa, American Journal of Mathematics Vol. 92, No. 4 (1970), pp. 997-1028.
\bibitem{Vidalcond} G. Vidal, J. I. Latorre, E. Rico, A. Kitaev, Phys. Rev. Lett. {\bf 90} 227902 (2003).
\bibitem{Wootters2} W. K. Wootters, Quant. Inf. Comp. Vol. 1, No. 1 27-44 (2001).
\bibitem{Schliemann} J. Schliemann, J. I. Cirac, M. KuS, M. Lewenstein and D. Loss, Phys. Rev. A {\bf 64} 022303 (2001).
\bibitem{Fierz} M. Fierz, Z. Physik {\bf 104} 553 (1937).
\bibitem{Fierz2} A. Miemiec, I. Schnakenburg, Fortsch. Phys. {\bf 54} 5-72 (2006).

\end{thebibliography}
\end{document}